\documentclass{article}
\usepackage[T1]{fontenc}
\usepackage[latin1]{inputenc}

\makeatletter

\providecommand{\LyX}{L\kern-.1667em\lower.25em\hbox{Y}\kern-.125emX\@}

 \newcommand{\lyxaddress}[1]{
   \par {\raggedright #1 
   \vspace{1.4em}
   \noindent\par}
 }

\usepackage[T1]{fontenc}
\usepackage[latin1]{inputenc}
\usepackage{geometry}


\begin{document}

\title{Interpolating Gauges, Parameter Differentiability ,WT-identities and the \( \epsilon  \)-term}

\author{Satish D. Joglekar\thanks{
e-mail address: sdj@iitk.ac.in 
}}

\maketitle

\lyxaddress{Department of Physics, I.I.T. Kanpur , Kanpur 208016 {[}India{]}}

\begin{abstract}
Evaluation of variation of a Green's function in a gauge field theory with a
gauge parameter \( \theta  \) involves field transformations that are (close
to) singular. Recently, we had demonstrated {[}hep-th/0106264{]} some unusual
results that follow from this fact for an interpolating gauge interpolating
between the Feynman and the Coulomb gauge (formulated by Doust). We carry out
further studies of this model. We study properties of simple loop integrals
involved in an interpolating gauge. We find several unusual features not normally
noticed in covariant Quantum field theories. We find that the proof of continuation
of a Green's function from the Feynman gauge to the Coulomb gauge \emph{via
such a gauge} in a gauge-invariant manner seems obstructed by the lack of differentiability
of the path-integral with respect to \( \theta  \) (at least at discrete values
for a specific Green's function considered) and/or by additional contributions
to the WT-identities. We show this by the consideration of simple loop diagrams
for a simple scattering process. The lack of differentiability, alternately,
produces a large change in the path-integral for a small enough change in \( \theta  \)
near some values. We find several applications of these observations in a gauge
field theory. We show that the usual procedure followed in the derivation of
the WT-identity that leads to the evaluation of a gauge variation of a Green's
function involves steps that are not always valid in the context of such interpolating
gauges. We further find new results related to the need for keeping the \( \epsilon  \)-term
in the in the derivation of the WT-identity and and a nontrivial contribution
to gauge variation from it. We also demonstrate how arguments using Wick rotation
cannot rid us of these problems. This work brings out the pitfalls in the use
of interpolating gauges in a clearer focus.\par{}
\end{abstract}

\section{Introduction}

{*}{*}{*}{*}{*}{*}{*}{*}{*}{*}{*}{*}{*}{*}{*}{*}{*}{*}{*}{*}{*}{*}{*}{*}{*}{*}{*}{*}{*}{*}{*}{*}{*}{*}{*}{*}{*}{*}{*}{*}{*}

The standard model \cite{cl} is a non-abelian gauge theory possessing a nonabelian
gauge-invariance. The consequences of the gauge-invariance have been formulated
as the WT-identities \cite{tv,lz} and are very important to the discussion
of renormalization and unitarity of the gauge theories\cite{iz,cl}. At the
practical as well as formal level, they are important in the discussion of gauge-independence
of observables. While the WT-identities in their usual form that is relevant
to the discussion of renormalizability ( i.e. structure of counterterms etc.)
are formulated via the usual (constant) BRS transformation, the form of the
WT-identities relevant to the discussion of gauge-independence is formulated
by considering a \emph{field-dependent} gauge transformation\cite{lz} or its
equivalent in the BRS formulation {[}These have been called IFBRS (Infinitesimal
Field-dependent BRS) transformations\cite{j}{]}. Recently, we found \cite{j}
reason to be cautious about the use of these field dependent transformations
as they are (close to) singular and have to be treated very carefully. We had
found, (and also given its justification), that while such a procedure that
uses these IFBRS transformations does not seem to lead to any obvious trouble
\emph{within the class} of the Lorentz-type gauges, it does indeed spell an
unexpected trouble for a class of interpolating gauges interpolating between
the Lorentz and the Coulomb gauge . We had found that in view of the {}``singular{}''
nature of the transformation involved, a careful treatment of the path-integral
\emph{including a correct \( \epsilon  \)-term} throughout was imperative and
we had further found that quite unexpected results follow from this treatment.
These results are further summarized in Sec. 2. (We expect this phenomenon to
be of a scope more general than the context in which it was analyzed). One of
the purposes of this work is to analyse these results in detail to shed new
light.

We shall now elaborate on the importance and scope of the subject matter discussed
in this work. Calculations in the standard model, a non-abelian gauge theory,
have been done in a variety of gauges depending on the ease and convenience
of calculations \cite{jijp,l,bns}. Many different gauges have also been used
in formal treatments in different contexts. For example axial gauges have been
used in the treatment of Chern-Simon theory, planar gauges in the perturbative
QCD, radial gauges in QCD sum-rules and Coulomb gauges in the confinement problem
in QCD \cite{refs}. Superstring theories also use to advantage both the covariant
and the light-cone treatments \cite{gsw}. One of the important questions, far
from obvious, has been whether the results for physical observables are, in
fact, independent of the choice of the gauge used in calculations. While it
has naively been assumed that this must do so, it is quite another matter to
actually prove the gauge independence of observables in general\footnote{%
As the path-integral in Lorentz-type gauges are well-defined, a little thought
will show that in the path-integral framework this is really a question of whether
and how path-integrals in other sets of gauges can be defined in a manner consistent
with the Lorentz gauges. 
}. A good deal of literature has been devoted to this question \cite{refs} directly
or indirectly.

The approach of interpolating gauges has been used to give a definition of gauges
other than the Lorentz gauges \cite{b,refs}. It has also been employed in formal
arguments that attempt to show the gauge-independence of observables in gauges
so connected to, say, the Lorentz gauges\cite{b,k}. The basic idea behind interpolating
gauges is to formulate the gauge theory in a gauge for which the gauge function
\( F[A(x);\alpha ] \) depends on one or more parameters \( \alpha  \) in such
a manner that for different values of the parameters we recover gauge theories
in different gauges. For example, the gauge function \( F[A(x);\theta ] \)
used by Doust \cite{d} to connect the Coulomb and the Feynman gauge is given
by\begin{equation}
\label{i.1}
F[A,\theta ]=[\theta \partial ^{0}A_{0}-\frac{1}{\theta }\partial _{i}A_{i}]
\end{equation}
 where for \( \theta  \)=1, we recover the Feynman gauge and for \( \theta  \)\( \rightarrow 0 \),
we recover the Coulomb gauge. Similarly, one could interpolate between the axial
and the Lorentz type gauges by a gauge function such as \begin{equation}
\label{i.2}
F[A,\kappa ,\lambda ]=\frac{1}{\sqrt{\lambda }}[(1-\kappa )\partial ^{\mu }A_{\mu }+\kappa \eta .A]
\end{equation}
 where for \( \kappa  \)=0, we recover a Lorentz-type gauge and for \( \kappa  \)=1,
we recover the axial gauge in the \( \lambda  \)\( \rightarrow 0 \) limit.
Such interpolating gauges have been employed in attempts to prove independence
of observables on the choice of the gauge. The arguments in such proofs proceed
along the same lines as those that prove the gauge independence of physical
observables under, say, a variation of the gauge parameter in the Lorentz type
gauges. It is here that we wish to first introduce a note of caution that while
such procedures may work within the class of Lorentz gauges, the results of
\cite{j} indicate that this may not be true for interpolating gauges from one
class to another as elaborated below. {[} The problem of definition of gauges
other than the Lorentz has in fact another solution that proceeds via a careful
way to link various pairs of gauges. This has already been introduced \cite{jold}
and results evaluated from these \cite{jmm} for various noncovariant gauges{]}.

In ref. \cite{j}, we had established several new observations regarding the
interpolating gauges. These observations apply to the type of the interpolating
gauges such as those given by (\ref{i.1}) {[}and possibly also to a large class
of similar interpolating gauges{]} but they do not affect the results while
dealing with the usual class of Lorentz-type gauges (in an unbroken theory at
least). These observations pertain to the role of the \( \epsilon  \)-term
in the path-integrals for such interpolating gauges. They are:

(i) While discussing the gauge variation, (e.g. varying \( \theta  \) in (\ref{i.1})),
the variation of the \( \epsilon  \)-term must be taken into account, if the
gauge-independence of the expectation value of a gauge-invariant operator is
to be at all preserved.

(ii) When this variation of the \( \epsilon  \)-term under \( \theta \rightarrow \theta +\delta \theta  \)
is taken into account, the net effect on the propagator is NOT an infinitesimal
one; but a drastic one if \( \frac{\delta \theta }{\varepsilon } \) is sufficiently
large\footnote{%
We note that \( \frac{\delta \theta }{\varepsilon } \)is a dimensionful quantity;
so the exact meaning of this qualitative characterrization involves other kinematical
quantities\cite{j}. 
}. These observations were further employed to imply that an interpolating gauge
as in (\ref{i.1}) \emph{with any simple \( \epsilon  \)-term} cannot interpolate
gauge-invariantly between two gauges.

While the above observations in \cite{j} could be interpreted only in a negative
light regarding the viability of using the interpolating gauges, the study of
interpolating gauges in \cite{j} also raises several unusual questions of a
\emph{general} nature about the derivation and the usage of the WT-identities
in the gauge theories especially when they concern the variation of Green's
functions with a gauge parameter. It is believed that these questions have enough
significance by themselves; independent of the context they arose from and could
have applications in gauge field theory in general. This article brings out
these questions into a clearer focus by a further study of the example in \cite{j}.
When we write down the WT-identities, we do not normally take account of the
variation of the \( \epsilon  \)-term in the path-integral. The results of
Ref. \cite{j} lead one to strongly suspect that this procedure to be a simplification
valid (probably) only for the class of the covariant Lorentz gauges. While dealing
with the \( \theta  \)-variation of the Green's functions in gauges such as
those in Eq. (\ref{i.1}), we find that we cannot drop the effect of the \( \epsilon  \)-term.
Further, even if we were to take into account the effect of the \( \theta  \)-variation
{[}\( \theta \rightarrow \theta +\delta \theta  \){]} of the \( \epsilon  \)-term
, observation (ii) above suggests that we cannot possibly regard the effect
as {}``infinitesimal{}'' if \( \frac{\delta \theta }{\varepsilon } \) is
sufficiently\footnote{%
See the earlier footnote regarding this qualitative characterization. 
} large \cite{j}.

These results point out to the fact that an infinitesimal variation in \( \theta  \)
{[}viz. \( \theta \rightarrow \theta +\delta \theta  \){]}, may sometimes produce
an {}``out of proportion{}'' effect in Green's functions in such formulations.
(In other words, the Green's functions may not be differentiable at some points).
In such cases the WT-identities derived by treating all variations as infinitesimal
and, in particular, ignoring the variation of the \( \epsilon  \)-term may
not always be the correct procedure.

In view of these observations in \cite{j}, we undertook to analyse the example
in \cite{j} in more details so that any unusual features of the gauge theories
in such gauges stand out. A close look at the analysis makes one believe that
the scope of the observations made here and in \cite{j} may be much more wide
than the specific context in which it was analysed.

In this work, therefore, we analyse in a greater detail some relevant simple
examples where there is reason to suspect new features not addressed to so far.
Using these, we aim to address the question of the usage of WT-identities in
the discussion of the gauge-dependence of Green's functions and observables
in the interpolating gauges.

In view of the unusualness of conclusions arrived at, we find it desirable to
analyze the issues in a fine detail so that there are no obvious loopholes.
We shall therefore first analyze a simple example by stages.

One of the essential points in \cite{j} was that as \( \theta \rightarrow \theta +\delta \theta  \),
the variation in the \( \epsilon  \)-term, even though infinitesimal of \( O[\varepsilon \delta \theta ] \)
formally, can in fact lead to a change in the propagator of much more significant
kind. This arose essentially from the fact that the effect of this term in some
kinematical region blows up. In this work, we want to analyze, in some greater
detail, effects of these kinds and further correlate these to the results of
Ref. \cite{j} and the expectations raised by it.

We now state the plan of the paper. In section 2, we shall introduce the notations
and summarize the conclusions of Ref. \cite{j}. In section 3, we shall pinpoint
the potential sources of trouble possibly requiring caution. We analyze it from
the point of view of the validity of the Taylor expansion of a propagator such
as \begin{equation}
\label{i.3}
D_{00}=\frac{1}{\theta ^{2}k_{0}^{2}-|\mathbf{k}|^{2}+i\varepsilon }
\end{equation}
 around \( \theta =\theta _{0} \) and the problems that it can lead to for
a k for which \( \theta ^{2}_{0}k^{2}_{0}-|\mathbf{k}|^{2}\approx 0 \). We
suspect two kinds of troubles: (i) One with differentiability in \( \theta  \)
as \( \epsilon \rightarrow 0 \). (ii) Second with order of the differentiation
\( \frac{\partial }{\partial \theta } \) and the limit \( \epsilon \rightarrow 0 \).

In section 4, we consider for illustration purposes, a simple model (noncovariant
\( \phi ^{4} \)- theory) where such a phenomenon can be analyzed in detail.
We, in fact, confirm the suspicions raised in the Sec 3. In section 5, we consider
the various orders of limits/ differentiations occurring in the definition of
gauge-variation of S-matrix and Green's functions. We analyze the procedure
normally used in this connection keeping in mind for contrast the observations
made in the section 4. Here, we point out that there are several delicate situations
in the derivation of WT identities and the discussion of gauge dependence. These
are the situations where (i) Order of limit \( \epsilon \rightarrow 0 \) and
differentiation with \( \theta  \), (ii) Expansion of an exponential with an
\char`\"{}infinitesimal\char`\"{} appearing exponent (iii) Keeping track of
the \( \epsilon  \)-term and its effects etc may have to be treated with care.

In section 6, we address to the question of the \( \theta  \)-dependence of
S-matrix elements and Green's functions in the interpolating gauge of Eq. (\ref{i.1}).
We focus our attention on a particular contribution to a simple 1-loop diagram.
We show that there exists a value of \( \theta \in (0,1) \) where this contribution
is not differentiable . What is worse, is that we find that the limits \( \epsilon \rightarrow 0 \)
and the differentiation do not commute at this point. In fact the right derivative
\( \frac{\partial }{\partial \theta }\Vert _{_{\theta ^{+}_{0}}} \)in fact
goes to infinity. Thus, it is generally not correct to assume that an infinitesimal
variation in \( \theta  \) will produce an infinitesimal variation in the path-integral
as is done in the derivation of a WT-identity. The simple example dealt with
here will also enable us to see that such situations will also exist in more
complicated loop diagrams. In section 7, we extend the above argument to a class
of off-shell Green's functions.

In Section 8, we shall analyse the results of section 6 in a direct manner.
One of the suspicions one may have had is that one should not have the problems
enumerated above because one could have formulated the field theory in Euclidean
space to begin with ; thus avoiding any necessity of an \( \epsilon  \)-term.
The analysis of this section will help us understand why this procedure cannot
rid us of the problems. Here, we find how the non-differentiability in \( \theta  \)
is connected to the presence of a branch-point in the complex \( s' \)-plane
(here, \( s'=\theta ^{2}p^{2}_{0}-|\mathbf{p}|^{2} \), see section 4 for more
on its definition). We also find out as to why these problem cannot be avoided
in the Euclidean formulation as they will pop up while carrying out the analytic
continuation from the Euclidean formulation.

In section 9, we derive a result which makes a contact between the result in
\cite{j} and the result in section 6 of this work. In section 10, we shall
examine the procedure we normally follow in the derivation of the WT-identities.
We point out, with reasons, the places that require a careful treatment. In
section 11, we add several comments. We comment on the extension of the result
about non-differentiability of the path-integral. We also comment on the use
of wavepckets for external lines. In appendix A, we shall present a simple example
of an off-shell Green's function where we explicitly confirm a nondifferentiable
behavior in \( \theta  \). In section 12, we summarise our conclusions.

This work brings out some of the pitfalls one has to face while formulating
a gauge-invariant field theory using interpolating gauges.

As a final comment, we note that despite the unusual nature of conclusions,
this work does not require more than a usual knowledge of Quantum Field Theory
and algebra.

\section{Summary of some results and Notations}

In this section, we shall summarize some results from the past works of references
\cite{d} and \cite{j} that are needed for our purpose. In the process, we
shall also introduce our notations.

In this work, we propose to discuss, in the path integral framework, some difficulties
encountered in the use of the interpolating gauges that interpolate between
pairs of gauges. Interpolating gauges are introduced by considering gauge functionals
that depend on one or more parameters. We shall, therefore, consider the Faddeev-Popov
effective action {[}FPEA{]} with a local gauge function \( F[A(x);\alpha ] \)
which may depend on several parameters, collectively denoted by \( \alpha  \)
. We denote this FPEA by \( S_{eff}[A,c,\overline{c};\alpha ] \) which is given
by \begin{equation}
\label{21.1}
S_{eff}[A,c,\overline{c};\alpha ]=S_{0}[A]+S_{gf}[A;\alpha ]+S_{gh}[A,c,\overline{c};\alpha ]
\end{equation}
 with\begin{equation}
\label{22.2}
S_{gf}[A;\alpha ]=-\frac{1}{2}\int d^{4}xF[A,\alpha ]^{2}
\end{equation}
 and\footnote{%
The ghost action is always arbitrary upto a constant and, in particular, an
overall sign. The following is a convention we make. 
}

\emph{\begin{equation}
\label{22.2b}
S_{gh}=-\int d^{4}xd^{4}y\overline{c}^{\alpha }(x)M^{\alpha \beta }[x,y;A;\alpha ]c^{\beta }(y)
\end{equation}
} with\begin{equation}
\label{21.3}
\int d^{4}yM^{\alpha \beta }[x,y;A;\alpha ]c^{\beta }(y)=\int d^{4}y\quad \frac{\delta F^{\alpha }[A(x);\alpha ]}{\delta A^{\gamma }_{\mu }(y)}D_{\mu }^{\gamma \beta }[A(y)]c^{\beta }(y)
\end{equation}
 and\begin{equation}
\label{21.3a}
D_{\mu }^{\alpha \beta }[A]=\delta ^{\alpha \beta }\partial _{\mu }+g\, f^{\alpha \beta \gamma }A_{\mu }^{\gamma }
\end{equation}
 Here, \( f^{\alpha \beta \gamma } \) are the antisymmetric structure constants
of a semi-simple gauge group.

\( S_{eff}[A,c,\overline{c};\alpha ] \) is invariant under the BRS transformations
:\begin{eqnarray}
\delta A^{\alpha }_{\mu }(x) & =D_{\mu }^{\alpha \beta }[A(x)]c^{\beta }(x)\delta \Lambda  & \nonumber \\
\delta c^{\alpha }(x) & =-\frac{1}{2}gf^{\alpha \beta \gamma }c^{\beta }(x)c^{\gamma }(x) & \delta \Lambda \nonumber \\
\delta \overline{c}^{\alpha }(x)=F[A(x);\alpha ]\delta \Lambda  & \label{21.4} 
\end{eqnarray}

In this work, we shall, in particular, utilize the example of the interpolating
gauge used by Doust to interpolate between the Feynman and the Coulomb gauge.
This example proves to be simple enough and yet bring out several problems associated
with these techniques, some of which were demonstrated in \cite{j}. It has
(as a special case of his)\begin{equation}
\label{21.5}
F[A,\theta ]=[\theta \partial ^{0}A_{0}-\frac{1}{\theta }\partial _{i}A_{i}]
\end{equation}
 We shall now summarize the results of \cite{j}. In this work we considered
the interpolating gauges, as for example those with an \( F \) given by (\ref{21.5}),
together with \emph{a conventional \( \epsilon  \)-term} viz. \begin{equation}
\label{21.5a}
-i\varepsilon \int d^{4}x[\frac{1}{2}A_{\mu }A^{\mu }-\overline{c}c]
\end{equation}
 {[}or a suitable modification that does not change \emph{signs} of terms in
(\ref{21.5a}){]} and discussed \emph{whether such a formulation can really
interpolate in a gauge-invariant way}. For this purpose, we considered the vacuum
expectation value of a gauge invariant operator \( O[A] \):\begin{equation}
\label{21.5b}
<O[A]>=\frac{<<O[A]>>}{<<1>>}
\end{equation}
 with\begin{eqnarray}
<<O[A]>>\mid _{_{\alpha _{0}}} & = & \int D\phi \, \, O[A]\exp \{iS_{0}[A]-\frac{i}{2}\int d^{4}xF[A(x),\alpha _{0}]^{2}+\nonumber \\
 &  & iS_{gh}[A,c,\overline{c};\alpha _{0}]+\int d^{4}x\varepsilon [\frac{1}{2}A^{2}-\overline{c}c]\}\label{21.6} 
\end{eqnarray}
 We then considered the question as to whether the above expression will necessarily
imply the independence of the expectation value with the parameter \( \alpha  \).
To see if this is so, we considered a field transformation of a infinitesimal
field-dependent BRS {[}IFBRS{]}-type (see e.g.\cite{j}; these are also spelt
out in section 10) that leads to\begin{eqnarray}
<<O[A]>>\mid _{_{\alpha _{0}}} & = & \int D\phi \, \, O[A]\exp \{iS_{0}[A]-\frac{i}{2}\int d^{4}xF[A(x),\alpha _{0}-\delta \alpha ]^{2}+\nonumber \\
 &  & iS_{gh}[A,c,\overline{c};\alpha _{0}-\delta \alpha ]+\int d^{4}x\varepsilon [\frac{1}{2}A^{2}-\overline{c}c]+\varepsilon \delta R\}\label{21.7} 
\end{eqnarray}
 We now note that the right hand side has an effective action evaluated at the
parameter \( \alpha _{0}-\delta \alpha  \); but at the same time the \( \epsilon  \)-term
has now changed it to\begin{equation}
\label{21.8}
\int d^{4}x\varepsilon [\frac{1}{2}A^{2}-\overline{c}c]\Longrightarrow \int d^{4}x\varepsilon [\frac{1}{2}A^{2}-\overline{c}c]+\varepsilon \delta R
\end{equation}
 with\begin{equation}
\label{21.9}
\delta R=-\varepsilon i\int d^{4}z\overline{c}\frac{\partial F}{\partial \alpha }\mid _{_{\alpha _{0}}}\int d^{4}x[\partial .A-F[A,\alpha ]c]\delta \alpha 
\end{equation}
 The effect of this term on the free propagator was then considered in the context
of the gauges of (\ref{21.5}) and next it was found that this term, rather
than having an infinitesimal effect on the propagator, in fact alters completely
the pole structure of the propagator for \( \delta  \)\( \alpha >>\varepsilon  \)\footnote{%
See the earlier footnote regarding this qualitative characterization. 
}. We then concluded that the small variation in \( \epsilon  \)-term with \( \theta  \)
has a catastrophic effect: it the gauge boson propagator structure from the
causal one to a mixed one even for a small change \( \delta  \)\( \theta  \).
We further demonstrated that there was no modification of the \( \epsilon  \)-term
that would allow us an escape in this gauge. We however found no such effect
from the \( \epsilon  \)-term for the set of Lorentz gauges as the gauge parameter
\( \lambda  \) is varied: the pole structure of the propagator does not alter
by such a term. This was verified for the unbroken gauge theory.

\section{Need for Caution in Field Theory Path-integrals}

\subsection{The origin of need for caution}

The path-integrals as used in perturbative quantum field theory are perturbations
over infinite dimensional Gaussian integrals. To begin with let us consider
a simple one-variable Gaussian integral:\begin{equation}
\label{2.1}
\int dx\exp \{iax^{2}-\varepsilon x^{2}\}
\end{equation}
 with 'a' real which, for \( \epsilon  \)> 0, as we know is\begin{equation}
\label{2.2}
\frac{1}{\sqrt{-i\pi (a-i\varepsilon )}}
\end{equation}
 For \( \epsilon  \) sufficiently small, we can expand this as \begin{equation}
\label{2.3}
\frac{1}{\sqrt{-i\pi a}}(1+\frac{i\varepsilon }{2a})
\end{equation}
 which holds provided \( \epsilon <a \) and may even ignore the \( \epsilon  \)-term
in (\ref{2.3}). We normally take for granted such approximations. In field
theory, we may come upon situations where \( a \) may be very small compared
to \( \epsilon  \) (or may even vanish) and there an approximation of this
kind may break down\footnote{%
At this point, it may be argued that a field theory can be formulated in the
Euclidean space where no epsilon term is requied. We shall discuss this question
in section 8 in further details. 
}. Moreover, there are situations where there may be no escape.

\subsection{A Field Theory Example}

Just to illustrate the above point in the context of a simple field theory first,
in a somewhat exaggerated fashion, we begin with a trivial observation in the
context of the \( \lambda  \)\( \phi  \)\( ^{4} \) -theory in the path-integral
formulation. We consider the generating functional \( W_{c}[J] \) for the Green's
functions in the Minkowski space:

\begin{equation}
\label{1.1}
W_{c}[J]=\int D\phi \exp \{iS_{0}[\phi ]-\varepsilon \int d^{4}x\phi ^{2}/2\}
\end{equation}
 where the suffix 'c' stands for the fact that \( W_{c}[J] \) generates the
causal Green's functions as ensured by the correct \( \epsilon  \)-term. We
may re-express:

\begin{equation}
\label{1.2}
W_{c}[J]=\int D\phi \exp \{iS_{0}[\phi ]+\varepsilon \int d^{4}x\phi ^{2}/2-2\varepsilon \int d^{4}x\phi ^{2}/2\}
\end{equation}
 We may, then, believe that the exponential \( \exp \{-\varepsilon \int d^{4}x\phi ^{2}\} \)
can be expanded in powers of the {}``small{}'' exponent \( \{-\varepsilon \int d^{4}x\phi ^{2}\} \)\footnote{%
Here, we ignore the possible subtleties that could be introduced by ultraviolet
divergences. We could for example stick to tree level. 
}.

\begin{eqnarray}
W_{c}[J] & = & \int D\phi \exp \{iS_{0}[\phi ]+\varepsilon \int d^{4}x\phi ^{2}/2\}[1-\varepsilon \int d^{4}x\phi ^{2}+.....]\label{1.3a} \\
 & \equiv  & W_{a.c.}[J]-\int D\phi \exp \{iS_{0}[\phi ]+\varepsilon \int d^{4}x\phi ^{2}/2\}\varepsilon \int d^{4}x\phi ^{2}+...]\label{1.3b} 
\end{eqnarray}
 where the suffix a.c. refers to {}``anti-causal{}''\footnote{%
We shall refer to the Green's functions with \( \epsilon \rightarrow -\varepsilon  \)
as the {}``anti-causal{}'' Green's functions. 
} Green's functions.

To see if such a procedure is \emph{always} a valid one, we consider the above
relation in the tree approximation where everything is expected to be well-defined.
We may then naively expect the second and higher terms in (\ref{1.3b}) to vanish
as \( \epsilon  \)\( \rightarrow  \)0. We would then obtain an absurd conclusion

\begin{equation}
\label{1.4}
\begin{array}{c}
lim\\
\varepsilon \rightarrow 0
\end{array}W_{c}[J]=\begin{array}{c}
lim\\
\varepsilon \rightarrow 0
\end{array}W_{a.c.}[J]
\end{equation}
 The above relation, in particular, implies for the propagator\begin{equation}
\label{1.5}
D_{c}(x-y)=D_{ac}(x-y)
\end{equation}
 which is incorrect since, for example, for \( x_{0}>y_{0}, \) the former propagates
+ve frequency modes and the latter the negative frequency modes.

The above example brings out the fact that such an exponential, with an exponent
{}``small{}'' in appearance, may not always be amenable to an expansion. This
happens, essentially because, the quadratic form in the path integral goes over
field configurations for which the {}``main term{}'' may in fact be small
compared to the {}``small term{}''; a situation very similar to Eq. (\ref{2.3})
being used under the wrong conditions \( a<\varepsilon  \).

To locate the mathematical error in the above argument in an alternate manner,
consider the 2-point function in the tree approximation generated by \( W_{c}[J] \)
of Eq. (\ref{1.3b}). In momentum space, it reads,

\begin{eqnarray}
\frac{1}{k^{2}-m^{2}+i\varepsilon } & = & \frac{1}{k^{2}-m^{2}-i\varepsilon +2i\varepsilon }\nonumber \\
 & = & \frac{1}{k^{2}-m^{2}-i\varepsilon }+\frac{-2i\varepsilon }{k^{2}-m^{2}-i\varepsilon }\frac{1}{k^{2}-m^{2}-i\varepsilon }+......\label{1.5a} 
\end{eqnarray}
 The equation (\ref{1.3b}) in the present context corresponds to this expansion.
We immediately recognize that the above Taylor expansion holds only if \( |2i\varepsilon |<|k^{2}-m^{2}-i\varepsilon | \).
Thus, the above procedure is not valid for a 4-dimensional volume in the \( k- \)space
\( \sim  \) \( \epsilon  \)\( \times  \) R\( ^{3} \).

Now the pertinent question is whether there is any place in field theoretic
calculations where such approximations are actually made and whether there are
cases where such points of mathematical rigor \emph{have} to be paid attention
to.

We shall show, in section 6, that while considering the gauge-parameter variation
of Green's functions in the interpolating gauges, we may have to pay special
attention to this point. There, we shall point out examples where an infinitesimal
change in a parameter leads to a disproportionate change in a Green's function.
In section 10, we shall also discuss the role of \( \epsilon  \)-terms (\( \varepsilon R \))
in the WT-identities. We had emphasized in Ref. \cite{j} that such terms may
have to be paid special attention to and we may not ignore effects arising from
them when a gauge parameter is varied {[}\( \epsilon R\rightarrow \varepsilon (R+\delta R) \){]}.
The effect was, in fact, such as to alter the boundary condition on the propagator.
We shall find that we may not be able to treat the exponents of a term of (\( \epsilon R \))
as amenable to expansion.

We know several cases where the propagator \emph{denominators} depend on a parameter.
For example, the gauge propagator in the R\( _{_{\xi }-} \)gauges \cite{fls}
has the form:

\begin{equation}
\label{1.6}
D_{\mu \nu }=\frac{g_{\mu \nu }-\frac{k_{\mu }k_{\nu }}{k^{2}-\frac{M^{2}}{\xi }+i\varepsilon }(1-\frac{1}{\xi })}{k^{2}-M^{2}+i\varepsilon }
\end{equation}
 and the associated ghost and the unphysical scalar fields have a similar \( \xi  \)-dependence
in the denominator. Similarly {[}a form of{]} an interpolating gauge that interpolates
between the Coulomb and the Feynman gauge used by Doust \cite{d} has the propagator
for (0,0) component\begin{equation}
\label{1.7}
D_{00}=\frac{1}{\theta ^{2}k_{0}^{2}-|\mathbf{k}|^{2}+i\varepsilon }
\end{equation}
 with a similar dependence in the spatial components. It is expected that the
various other interpolating schemes \cite{b,bz} will also have propagator denominators
depending on interpolating parameters in such sensitive a manner.

A typical Feynman diagram, in the context of the above example (28), depends
on \( \theta  \) through such propagators. While considering the parameter
variation of a Feynman diagram with the gauge parameter \( \theta  \), we are
in effect expanding each propagator in a Taylor series around \( \theta  \)\( _{0} \)
in a series such as

\begin{eqnarray}
D_{00} & = & \frac{1}{\theta ^{2}k_{0}^{2}-|\mathbf{k}|^{2}+i\varepsilon }\\
 & = & \frac{1}{\theta _{0}^{2}k_{0}^{2}-|\mathbf{k}|^{2}+i\varepsilon }+\frac{1}{\theta _{0}^{2}k_{0}^{2}-|\mathbf{k}|^{2}+i\varepsilon }\bullet \frac{-2\theta _{0}\delta \theta k^{2}_{0}}{\theta _{0}^{2}k_{0}^{2}-|\mathbf{k}|^{2}+i\varepsilon }+......\label{1.8} 
\end{eqnarray}
 and picking the second term on the right side to compute the \( \theta  \)-
variation. We, however, note that this Taylor series is valid only if \begin{equation}
\label{1.9}
|\frac{-2\theta _{0}\delta \theta k^{2}_{0}}{\theta _{0}^{2}k_{0}^{2}-|\mathbf{k}|^{2}+i\varepsilon }|<1
\end{equation}

We, thus, note several points:

{[}a{]} We have to be careful while using the above Taylor series (or a procedure
equivalent to it) in formal manipulations since it may not always hold for arbitrary
\( \delta  \)\( \theta  \);

{[}b{]} For \( \delta \theta  \) sufficiently large {[}with some small fixed
\( \epsilon  \){]} , the above condition (\ref{1.9}) is not valid for a 4-dimensional
volume \( \sim \delta  \)\( \theta  \) X R\( ^{3} \) in the \( k \)-space.

{[}c{]} For a fixed \( \epsilon >0 \), and a finite \( k_{0} \) , there is
a nonzero range of \( \delta \theta  \) for which the above condition is necessarily
fulfilled. \emph{Nonetheless, this range \( \rightarrow 0 \) , as \( \epsilon  \)\( \rightarrow 0 \).
Recall that to evaluate the S-matrix elements, we always take the limit \( \epsilon  \)\( \rightarrow 0 \).}

{[}d{]} We note in passing that in the 4-dimensional subspace: \( \theta  \)\( ^{2}_{_{0}}k^{2}_{0} \)-|\textbf{k|\( ^{2} \)}\( \simeq  \)0,
the above ratio in Eq. (\ref{1.9}) is \( \sim \frac{\delta \theta }{\varepsilon } \).
And this is precisely the parameter on which an unexpected dependence was found
\cite{j} in the discussion of the gauge variation of a path-integral in this
sort of an interpolating gauge. We also note that for such a \( k \), the radius
of convergence of the series (\ref{1.8}) is \( \sim \frac{\delta \theta }{\varepsilon } \).
This is a simple and clear illustration of how the two independent small parameters
\( \delta  \)\( \theta  \) and \( \epsilon  \) become entangled!

{[}e{]} The naive expectation that \( \epsilon  \) is just a spectator in the
entire discussions of WT-identities and of gauge-independence in gauge field
theories and it need not be paid special attention to does not always seem valid.
According to the results of \cite{j}, this is the case, probably only for the
class of the Lorentz gauges.

{[}f{]} The point {[}c{]} above suggests the possibility that there could arise
a difficulty in the definition of a derivative with respect to \( \theta  \)
in loop integrals as \( \epsilon \rightarrow 0 \) and that the behavior of
loop integrals could possibly depend on the orders of \( \frac{d}{d\theta } \)
and the limits \( \epsilon  \)\( \rightarrow  \)0. We shall confirm these
suspicions in Sections 4 and 6.

At this point, one may wonder whether working with a Wick-rotated Euclidean
field theory will not rid us of all such problems as then \( \epsilon  \) would
be redundant. We shall clarify this point in the Section 8. It turns out that
this is not always possible.

\section{Direct analysis of nondifferentiability of a generating functional}

The purpose of this section is to make clear, in a direct fashion, the difficulty
in defining the partial derivative of a generating functional {[}see e.g. eq.
(\ref{abc}) below{]} with respect a parameter \( \alpha  \), on which it depends,
when the dependence on \( \alpha  \) is such that a propagator denominator
depends sensitively on it in some domain in the momentum space. As expected
from the previous section, this happens when the propagator singularities depend
on \( \alpha  \).

We shall find it useful to illustrate the point by considering an artificial
but a simple model that captures, in a direct manner, the essential point made
in the previous section. Rather than get into complications of a generating
functional of a gauge theory in the beginning itself, we shall consider a scalar
field theory with a specific (but noncovariant\emph{)} kinetic energy term\emph{.}
The generating functional of Green's functions is given by\begin{eqnarray}
W[J,\alpha ] & =\begin{array}{c}
lim\\
\varepsilon \rightarrow 0
\end{array}W[J,\alpha ,\varepsilon ] & \\
= & \begin{array}{c}
lim\\
\varepsilon \rightarrow 0
\end{array}\int D\phi \exp \{i\int d^{4}x[\frac{1}{2}\phi (\alpha \partial ^{2}_{0}-\nabla ^{2}-m^{2}+i\varepsilon )\phi +\frac{\lambda }{4}\phi ^{4}]+i\int d^{4}xJ\phi \};\, \alpha >0\label{abc} 
\end{eqnarray}
 We thus note that to compute \( \frac{\partial W[J,\alpha ]}{\partial \alpha } \),
we must first evaluate

\begin{equation}
\label{5.2}
\begin{array}{c}
lim\\
\varepsilon \rightarrow 0
\end{array}\{W[J,\alpha +\delta \alpha ,\varepsilon ]-W[J,\alpha ,\varepsilon ]\}
\end{equation}
 which requires the evaluation of \begin{equation}
\label{5.3}
\begin{array}{c}
lim\\
\varepsilon \rightarrow 0
\end{array}\int D\phi \exp \{i\int d^{4}x[\frac{1}{2}\phi (\alpha \partial ^{2}_{0}-\nabla ^{2}-m^{2}+i\varepsilon )\phi +\frac{\lambda }{4}\phi ^{4}]+i\int d^{4}xJ\phi \}\bullet \left( \exp [i\delta \alpha \int d^{4}x\phi \partial ^{2}_{0}\phi ]-1\right) 
\end{equation}
 We wish to pose the following question: can one, for \emph{any} small enough
\( \delta  \)\( \alpha  \), \emph{and irrespective of \( \epsilon  \)}, expand
the exponential, \emph{within the path integral}, as

\begin{equation}
\label{5.4}
\exp [i\delta \alpha \int d^{4}x\phi \partial ^{2}_{0}\phi ]\approx 1+O[\delta \alpha ]
\end{equation}

To see this, we first consider the bare two point function of the expression
(\ref{5.3}), in the momentum space:\begin{equation}
\label{5.5}
\frac{1}{(\alpha +\delta \alpha )k_{0}^{2}-|\mathbf{k}|^{2}-m^{2}+i\varepsilon }-\frac{1}{\alpha k_{0}^{2}-|\mathbf{k}|^{2}-m^{2}+i\varepsilon }
\end{equation}
 We note that we can generally Taylor expand the first term as

\begin{equation}
\label{5.6}
\frac{1}{\alpha k_{0}^{2}-|\mathbf{k}|^{2}-m^{2}+i\varepsilon }+\frac{1}{\alpha k_{0}^{2}-|\mathbf{k}|^{2}-m^{2}+i\varepsilon }\bullet \frac{-\delta \alpha k^{2}_{0}}{\alpha k_{0}^{2}-|\mathbf{k}|^{2}-m^{2}+i\varepsilon }+......
\end{equation}
 provided \( |\alpha k^{2}_{0}-|\mathbf{k}|^{2}-m^{2}+i\varepsilon |>|\delta \alpha k^{2}_{0}| \).
We now recall the discussion in the previous section and note that, evidently,
the magnitude of \( \epsilon  \) comes into play in being able to carry out
the expansion when \( k \) is such that \( \alpha k^{2}_{0}-|\mathbf{k}|^{2}-m^{2}\simeq 0 \)!

Now, the fact that \emph{the Taylor expansion of (\ref{5.6}) fails as \( \epsilon  \)\( \rightarrow 0 \)
in a subspace does seem to have an effect in a one-loop diagram}. To illustrate
this effect, we wish to consider an \( s- \)channel\footnote{%
We shall be interested later in the \emph{imaginary} part of the amplitude.
Only the s-channel diagram can have this imaginary part for the physical amplitude. 
} diagram for the process \( \phi \phi \rightarrow \phi \phi  \) in the one
loop approximation. This process involves the integral\footnote{%
This work involves noncovariant formulations. Thus several results will necessarily
be explicitly Lorentz frame dependent. 
}\begin{equation}
\label{5.7}
I(p,m,\alpha ,\varepsilon )=i\int \frac{d^{4}k}{(\alpha k^{2}_{0}-|\mathbf{k}|^{2}-m^{2}+i\varepsilon )[\alpha (k+p)^{2}_{0}-|\mathbf{k}+\mathbf{p}|^{2}-m^{2}+i\varepsilon ]}
\end{equation}
 We shall find it convenient to relate this to the {}``fish diagram{}'' in
the usual covariant \( \phi ^{4} \)-theory \cite{r}. We define,\begin{equation}
\label{5.7a}
p_{\mu }'=(\sqrt{\alpha }p_{0},\mathbf{p});\qquad k_{\mu }'=(\sqrt{\alpha }k_{0},\mathbf{k})
\end{equation}
 so that, we have\begin{equation}
\label{5.8}
I(p,m,\alpha ,\varepsilon )=\frac{1}{\sqrt{\alpha }}F(p',m,\varepsilon )
\end{equation}
 where \( F \) stands for the {}``fish diagram{}'' amplitude in the \( \phi  \)\( ^{4} \)-theory:\begin{equation}
\label{5.9}
F(p,m,\varepsilon )=i\int \frac{d^{4}k}{(k^{2}_{0}-|\mathbf{k}|^{2}-m^{2}+i\varepsilon )[(k+p)^{2}_{0}-|\mathbf{k}+\mathbf{p}|^{2}-m^{2}+i\varepsilon ]}
\end{equation}

Now, we know how this is evaluated and know the analytic properties of the amplitude
as seen from any text-book \cite{r} .

We define \( s=p^{2} \) and \( s'=p'^{2} \). \( F \) can be looked upon as
a function of a complex variable \( s=p^{2} \). The analytic properties of
\( F(p,m,\varepsilon ) \) on the real-\( s \) axis depend on whether \( s>4m^{2} \)
or \( s<4m^{2} \). Thus, the analytic properties of \( I(p,m,\alpha ,\varepsilon ) \)
will depend on whether \( s'=\alpha p^{2}_{0}-|\mathbf{p}|^{2}>4m^{2} \) or
\( s'<4m^{2} \). For \( s'<4m^{2} \), we can perform a Wick rotation and evaluate
the integral, by say, a cutoff method\footnote{%
Later, we shall be focussing on the imaginary part, which is finite.Hence the
details of regularization does not matter. 
}. The result is {[} here, A is a real constant{]}:\begin{eqnarray}
I(p,m,\alpha ,\varepsilon )= & \frac{1}{\sqrt{\alpha }}A\int ^{1}_{0}dx\ln \frac{[m^{2}-x(1-x)s'-i\varepsilon ]}{\Lambda ^{2}} & \label{5.11} \\
= & \frac{1}{\sqrt{\alpha }}A\int ^{1}_{0}dx\ln \frac{[m^{2}-x(1-x)(\alpha p^{2}_{0}-|\mathbf{p}|^{2})-i\varepsilon ]}{\Lambda ^{2}}\label{5.12} 
\end{eqnarray}

Now for \( \Re e\, s'<4m^{2} \), this quantity is analytic, and it has a branch-cut
from \( s'=4m^{2} \) along the real \( s' \) axis {[}actually, slightly below
the real axis{]}. We, now, will drop the real term \( \sim \ln \Lambda ^{2} \)
and focus attention on the finite part \( \frac{A}{\sqrt{\alpha }}I'(p,m,\alpha ,\varepsilon ) \).
We note that the logarithm has a phase varying from -\( \pi  \) to \( \pi  \)
. We then have,\begin{eqnarray}
I'(p,m,\alpha ,\varepsilon ) & = & \int ^{1}_{0}dx\frac{1}{2}\ln \{[m^{2}-x(1-x)(\alpha p^{2}_{0}-|\mathbf{p}|^{2})]^{2}+\varepsilon ^{2}\}\nonumber \\
 &  & +\int ^{1}_{0}dx[-\pi i+i\arctan \{\frac{\varepsilon }{x(1-x)(\alpha p^{2}_{0}-|\mathbf{p}|^{2})-m^{2}}\}]\label{5.13} 
\end{eqnarray}
 Here, we have chosen the range of \( \arctan  \) as {[}0,\( \pi  \)) for
our convenience. {[}Note that this definition differs from the principal branch
\( \arctan  \){]}. We note, in particular that \begin{equation}
\label{5.13a}
\begin{array}{c}
lim\\
\varepsilon \rightarrow 0
\end{array}ImI'(p,m,\alpha ,\varepsilon )=\int ^{1}_{0}dx\{-\pi i\Theta [B]\}=-\pi i\Delta x
\end{equation}
 with \( \Theta  \){[}B{]} as the step function and\begin{equation}
\label{5.13b}
B=x(1-x)(\alpha p^{2}_{0}-|\mathbf{p}|^{2})-m^{2}=x(1-x)s'-m^{2}
\end{equation}
 and \( \Delta x \) is the interval over which \( B>0 \). Now, for the s-channel
process \( \phi \phi \rightarrow \phi \phi  \), we evaluate the expression
of (\ref{5.3}) in the one loop approximation. We, then, have for the difference,

\begin{eqnarray}
 &  & Im\left\{ \begin{array}{c}
lim\\
\varepsilon \rightarrow 0
\end{array}I'(\alpha +\delta \alpha ,p,m,\varepsilon )-\begin{array}{c}
lim\\
\varepsilon \rightarrow 0
\end{array}I'(\alpha ,p,m,\varepsilon )\right\} \nonumber \\
 & = & Im\left\{ \Pi _{i}(p^{2}_{i}-m^{2}_{i})FT\frac{\delta ^{4}}{\delta J^{4}}\begin{array}{c}
lim\\
\varepsilon \rightarrow 0
\end{array}<<\exp [i\delta \alpha \int d^{4}x\phi \partial ^{2}_{0}\phi ]-1>>\mid _{_{J=0}}\right\} \label{5.14} 
\end{eqnarray}
 where, \( \frac{\delta ^{4}}{\delta J^{4}} \)is a brief way to express the
act of extracting the 4-point function, \( FT \) stands for the Fourier transform.
We now choose \( \alpha  \)=\( \alpha  \)\( _{0} \) such that \( s'=4m^{2} \)and
choose a (\( \alpha _{0}+\delta \alpha ) \) such that \( s'>4m^{2} \). We
now take the imaginary part of both sides. We note:

\subparagraph{\textmd{(i) I'(\protect\protect\( \alpha _{0}\protect \protect \)+\protect\protect\( \delta \protect \protect \)\protect\protect\( \alpha \protect \protect \))
has a nonzero imaginary part as \protect\protect\( \epsilon \protect \protect \)\protect\protect\( \rightarrow \protect \protect \)0.} }

\subparagraph{(ii)\textmd{I'(\protect\protect\( \alpha ^{-}_{0}\protect \protect \)) has}
\textmd{\emph{no}} \textmd{imaginary part as \protect\protect\( \epsilon \protect \protect \)\protect\protect\( \rightarrow \protect \protect \)0.}
\textmd{Here, \protect\protect\( \alpha _{0}^{-}\protect \protect \)stands
for the limit from left \protect\protect\( \alpha \rightarrow \alpha _{0}\protect \protect \).
This is so since \protect\protect\( B\protect \protect \) of (\ref{5.13b})
is necessarily negative for \protect\protect\( \alpha <\alpha _{0}\protect \protect \).}}

\subparagraph{\textmd{(iii)The difference of the imaginary parts is} \textmd{\emph{not}}
\textmd{proportional to \protect\protect\( \delta \alpha \protect \protect \);
a simple calculation shows that for small \protect\protect\( \delta \alpha \protect \protect \),
it is proportional to \protect\protect\( \sqrt{\delta \alpha }\protect \protect \).
To see this, we note that with \protect\protect\( \alpha =\alpha _{0}+\delta \alpha \protect \protect \),
\protect\protect\( s'=4m^{2}+\delta \alpha p^{2}_{0}\protect \protect \) ;
and for \protect\protect\( x\sim \frac{1}{2}\protect \protect \)}}

\subparagraph{\textmd{\protect\protect\( B=-4m^{2}(x-\frac{1}{2})^{2}+x(1-x)\delta \alpha p^{2}_{0}\simeq -4m^{2}(\Delta x)^{2}+\frac{1}{4}\delta \alpha p^{2}_{0}\protect \protect \)}}

and thus, \( B>0 \) over a range if \( x \), viz. \( 2\Delta x\sim \sqrt{\delta \alpha } \).

\subparagraph{\textmd{(iv) On account of (iii), the difference of the imaginary parts cannot
be understood as}}

\begin{equation}
\label{5.15}
Im\Pi _{i}(p^{2}_{i}-m^{2}_{i})FT\frac{\delta ^{4}}{\delta J^{4}}\begin{array}{c}
lim\\
\varepsilon \rightarrow 0
\end{array}<<[i\delta \alpha \int d^{4}x\phi \partial ^{2}_{0}\phi ]>>\mid _{_{J=0}}
\end{equation}
 which is proportional to \( \delta \alpha  \).

(v)In fact, a simple analysis will reveal that

\begin{equation}
\label{5.16}
\Pi _{i}(p^{2}_{i}-m^{2}_{i})FT\frac{\delta ^{4}}{\delta J^{4}}\begin{array}{c}
lim\\
\varepsilon \rightarrow 0
\end{array}<<[i\delta \alpha \int d^{4}x\phi \partial ^{2}_{0}\phi ]>>\mid _{_{J=0;\alpha \rightarrow \alpha ^{-}_{0}}}
\end{equation}
 does not have an imaginary part. This is seen as follows: The expression (\ref{5.16})
is understood as being proportional to \( \delta \alpha \frac{\partial }{\partial \alpha }\begin{array}{c}
lim\\
\varepsilon \rightarrow 0
\end{array}I(p,m,\alpha ,\varepsilon )\mid _{_{\alpha ^{-}_{0}}} \) which is real.

(vi) We further note that as a result of (iii),

\[
\frac{\partial }{\partial \alpha }\begin{array}{c}
lim\\
\varepsilon \rightarrow 0
\end{array}I'(\alpha ,p,m,\varepsilon )\]
 is ill-defined at this value of \( \alpha  \). We shall discuss this in much
more detail in section 6 where we will employ discussion carried out in this
section in the context of a gauge field theory 1-loop calculation.

Finally, we add a comment on whether we could avoid such problems if we do the
field theory keeping \( \epsilon  \) small but nonzero. We then note that derivative
of \( I \) with \( \alpha  \) will exist, and so will the Taylor series of
I around \( \alpha =\alpha _{0} \). But now the derivative will be a very large
number \( \rightarrow \infty  \) as \( \epsilon \rightarrow 0 \); and the
radius of convergence of the Taylor expansion will be very small.So we may not
calculate, by a Taylor expansion, the value of \( I(\alpha _{0}+\delta \alpha ) \)
from that for \( I(\alpha _{0}) \) if {}``\( \delta \alpha >>\varepsilon  \){}''
. Moreover, we should note that the limit \( \epsilon \rightarrow 0 \) is required
for a physical (unitary) theory because even in a \( \lambda  \)\( \phi  \)\( ^{4} \)-theory
with \( L'=L+i\varepsilon \phi ^{2} \), a Hermitian Hamiltonian ( and a unitary
S-matrix) is obtained only as \( \epsilon \rightarrow 0 \).

We also note the reason we consider this question at all. In a gauge field theory,
we derive the WT-identity for gauge variation, we need to confirm whether such
variations will be infinitesimal of first order always.We shall find that this
is not always so in the interpolating gauges and we find some unexpected results
from it in sections 9-10.

\section{Delicate limits in the discussion of gauge-independence}

In the previous sections, we noted places in Quantum field theory calculations
where care may me needed. We wish now to note down the procedure that we normally
follow for establishing the WT-identity used in evaluating gauge variation of
Green's functions. We do this with a view to see if any of the steps could fail
in gauges such those under consideration. We shall then discuss several points
regarding the derivation in Sections 6-11.

We consider the generating functional of the unrenormalized connected Green's
functions \( Z[J,\theta ,\varepsilon ] \). We obtain the (renormalized) S-matrix
elements from it by the operation \( \Theta  \): defined as a succession of
the following:\begin{equation}
\label{4.1}
S(p_{i},\theta )=\Theta Z[J,\theta ,\varepsilon ]\equiv \begin{array}{c}
lim\\
\varepsilon \rightarrow 0
\end{array}S(p_{i},\theta ,\varepsilon )
\end{equation}

\begin{equation}
\label{4.2}
\Theta \equiv \begin{array}{c}
lim\\
\varepsilon \rightarrow 0
\end{array}\begin{array}{c}
lim\\
p^{2}_{i}\rightarrow m^{2}_{i}-i\varepsilon 
\end{array}\Pi _{i}(p^{2}_{i}-m^{2}_{i}+i\varepsilon )\frac{\phi _{i}(p_{i})}{\sqrt{Z_{i}}}F.T.[\frac{\delta ^{n}}{\delta J^{n}(x)}]
\end{equation}
 where the \( n^{th} \) order functional derivative acting on \( Z[J,\theta ,\varepsilon ] \)
at \( J=0 \) is written as \( [\frac{\delta ^{n}}{\delta J^{n}(x)}] \) for
brevity; \( F.T. \) stands for the Fourier transform; \( Z_{i}(\theta ) \)'s
stand for the mass-shell renormalization constants and \( \phi _{i}(p_{i}) \)
is the physical wavefunction\footnote{%
We note the somewhat unconventional appearance of truncation factor \( (p^{2}_{i}-m^{2}_{i}+i\varepsilon ) \)
and the on-shell limit \( p^{2}_{i}\rightarrow m^{2}_{i}-i\varepsilon  \).
Before we let \( \varepsilon \rightarrow 0 \), we have to do this to correctly
truncate the external line propagators. Also, in the discussion of gauge-independence,
we need to remove any pole-less contributions to the \( \frac{\partial }{\partial \theta }G \);
and these are then correctly removed by the above on-shell limit \( p^{2}_{i}\rightarrow m^{2}_{i}-i\varepsilon  \).
The parameter \( \epsilon  \) is let go to zero only in the end. 
}. The gauge independence of \( S(p_{i},\theta ,\varepsilon ) \) is expressed
by the requirement \begin{equation}
\label{4.3}
\frac{\partial }{\partial \theta }S(p_{i},\theta )=\frac{\partial }{\partial \theta }\begin{array}{c}
lim\\
\varepsilon \rightarrow 0
\end{array}S(p_{i},\theta ,\varepsilon )=0
\end{equation}
 It is to be particularly noted that the order of the limit (\( \epsilon  \)\( \rightarrow 0) \)
and differentiation \emph{\( \frac{\partial }{\partial \theta } \)} is as follows:
\( \frac{\partial }{\partial \theta }\begin{array}{c}
lim\\
\varepsilon \rightarrow 0
\end{array} \)

We shall, now, recapitulate how one actually obtains the gauge-dependence of
the on-shell physical Green's functions in practice:

\subparagraph{\textmd{(a)We start from \protect\protect\( Z[J,\theta ,\varepsilon ]\protect \protect \)
and evaluate \protect\protect\( \frac{\partial }{\partial \theta }\protect \protect \)\protect\protect\( Z[J,\theta ,\varepsilon ]\protect \protect \)
for a fixed \protect\protect\( \epsilon \protect \protect \). }}

\subparagraph{\textmd{(b)We then use the WT identities to simplify the quantity under question
\cite{lz}.}}

\subparagraph{\textmd{(c)We find the part of this that contributes to the} \textmd{\emph{on-shell
physical}} \textmd{Green's functions. In other words, we evaluate}}

\begin{equation}
\label{4.4}
\Theta \frac{\partial }{\partial \theta }Z[J,\theta ,\varepsilon ]=\begin{array}{c}
lim\\
\varepsilon \rightarrow 0
\end{array}\begin{array}{c}
lim\\
p^{2}_{i}\rightarrow m^{2}_{i}-i\varepsilon 
\end{array}\Pi _{i}(p^{2}_{i}-m^{2}_{i}+i\varepsilon )\frac{\partial }{\partial \theta }\frac{\phi _{i}(p_{i})}{\sqrt{Z_{i}}}F.T.[\frac{\delta ^{n}}{\delta J^{n}(x)}]Z[J,\theta ,\varepsilon ]
\end{equation}

We wish to note several points in the above derivation:

{[}1{]} We first assume that the limit \( \epsilon  \)\( \rightarrow  \)0
and the differentiation \( \frac{\partial }{\partial \theta } \) can be interchanged.

{[}2{]} We evaluate then \( \frac{\partial W[J,\theta ,\varepsilon ]}{\partial \theta } \)by

\begin{equation}
\label{4.5}
\frac{\partial W[J,\theta ,\varepsilon ]}{\partial \theta }=\frac{\partial }{\partial \theta }\int D\phi \exp \left( iS_{eff}[\phi ,\theta ]+i\int d^{4}xJ\phi \right) 
\end{equation}
 . In doing so, it is being tacitly assumed that this derivative always exists
even as \( \epsilon \rightarrow 0 \). We have already seen a counter-example\footnote{%
At this point, it may be thought that it is not important whether the derivative
\( \frac{dI}{d\theta } \)exists at a point \( \theta _{0} \); it is only required
that \( \int \frac{dI}{d\theta }d\theta  \) exists over a small interval covering
\( \theta _{0} \). After all, we may make \( \frac{dI}{d\theta } \) finite
by keeping \( \epsilon  \) finite till end (or possibly by using suitable wavepackets
for external lines: see section 11). But this does not evade the problem in
section 10 about the contribution of the \( \epsilon  \)-term. We find the
language of {}``non-differentiability{}'' as the best way to exhibit the problem. 
} in Sec.4 and we shall examine this question in Section 6 further.

{[}3{]} Equivalently, the derivative is evaluated by the assumption that\begin{equation}
\label{4.6}
\frac{\partial }{\partial \theta }\int D\phi \exp \left( iS_{eff}[\phi ,\theta ]+i\int d^{4}xJ\phi \right) 
\end{equation}

can be evaluated by expressing it as \begin{equation}
\label{4.7}
i\int D\phi \frac{\partial }{\partial \theta }S_{eff}[\phi ,\theta ]\exp \left( iS_{eff}[\phi ,\theta ]+i\int d^{4}xJ\phi \right) 
\end{equation}

In this we make an assumption that \begin{equation}
\label{4.9}
\int D\phi \exp \left( iS_{eff}[\phi ,\theta ]+i\int d^{4}xJ\phi \right) \left\{ \exp (iS_{eff}[\phi ,\theta +\delta \theta ])-\exp (iS_{eff}[\phi ,\theta ])\right\} 
\end{equation}
 can always be written as \begin{equation}
\label{4.10}
i\int D\phi \frac{\partial }{\partial \theta }S_{eff}[\phi ,\theta ]\delta \theta \exp \left( iS_{eff}[\phi ,\theta ]+i\int d^{4}xJ\phi \right) 
\end{equation}
 even as \( \epsilon \rightarrow 0 \). In this connection, we draw attention
to the example in the section 4.

{[}4{]} In writing down the WT-identities, we do not keep track of the contribution
to it from the \( \epsilon  \)-term. We recall that recently, we have noted
that \cite{j} a careful attention to the \( \epsilon  \)-term has to be given
in the case of gauges other than the Lorentz gauges.

{[}5{]} We then simplify the resultant expression by the use of the WT-identities.
We then pick out the part of this that contributes to the truncated physical
Green's functions.

We shall analyze these points in detail in the next section.

\section{An Explicit Example}

In this section, we shall make a number of obervations that have a potential
bearing on the derivation of WT identities for evaluating gauge-dependence of
Green's functions.

We start with the example of the gauge theory in an interpolating gauge of Eq.
(\ref{21.5}) so defined as to interpolate between the Feynman and the Coulomb
gauge \cite{d}. For simplicity, we shall first focus our attention on a particular
integral\footnote{%
It is of course true that very often the \emph{total} amplitude in a gauge theory
has properties different from a specific piece of it; this being due to cancellations
of terms. 
} that appears in a contribution to the 4-point function (of gluons) in the one
loop approximation. It appears in the one loop diagram with two 4-point vertices
and 2 \( A_{0} \) internal lines\footnote{%
We are focussing attention on a part of the contribution due to time-like gluons. 
}. The integral reads\begin{equation}
\label{6.1}
I(p,\theta ,\varepsilon )=i\int \frac{d^{4}k}{(\theta ^{2}k^{2}_{0}-|\mathbf{k}|^{2}+i\varepsilon )[\theta ^{2}(k+p)^{2}_{0}-|\mathbf{k}+\mathbf{p}|^{2}+i\varepsilon ]}
\end{equation}
 To correlate the above integral with with that of Section 4 {[}See eq. (\ref{5.7}){]},
we shall find it convenient to rather consider an integral with a \emph{mass}
included\footnote{%
We could regard this as an infrared regularization of the amplitude in question.
We shall, however, eventually focus on the imaginary part of (\ref{6.1}) and
this is infrared finite here. 
}. We consider,\begin{equation}
\label{6.2}
I(p,m,\theta ,\varepsilon )=i\int \frac{d^{4}k}{(\theta ^{2}k^{2}_{0}-|\mathbf{k}|^{2}-m^{2}+i\varepsilon )[\theta ^{2}(k+p)^{2}_{0}-|\mathbf{k}+\mathbf{p}|^{2}-m^{2}+i\varepsilon ]}
\end{equation}
 We shall then explicitly demonstrate that, in the physical region, i.e. for
\( p^{2}>4m^{2} \), there exists a value of \( \theta  \) viz. \( \theta  \)\( _{0} \)\( \in  \){[}0,1{]}
such that at this point

(a) The derivative \( \frac{\partial }{\partial \theta } \)\( \begin{array}{c}
lim\\
\varepsilon \rightarrow 0
\end{array} \)\( I(p,m,\theta ,\varepsilon ) \) does not exist: i.e.\begin{equation}
\label{6.2a}
\frac{\partial }{\partial \theta }\begin{array}{c}
lim\\
\varepsilon \rightarrow 0
\end{array}I(p,m,\theta ,\varepsilon )\mid _{_{\theta ^{-}_{0}}}\neq \frac{\partial }{\partial \theta }\begin{array}{c}
lim\\
\varepsilon \rightarrow 0
\end{array}I(p,m,\theta ,\varepsilon )\mid _{_{_{\theta _{0}^{+}}}}
\end{equation}

(b) What is worse is that the imaginary part of the left hand side vanishes
at this point, whereas the imaginary part of the right hand side \( \rightarrow  \)
infinity. Thus, this quantity has infinite discontinuity.

(c)Moreover, we find that\[
\begin{array}{c}
lim\\
\varepsilon \rightarrow 0
\end{array}\frac{\partial }{\partial \theta }I(p,m,\theta ,\varepsilon )-\frac{\partial }{\partial \theta }\begin{array}{c}
lim\\
\varepsilon \rightarrow 0
\end{array}I(p,m,\theta ,\varepsilon )\mid _{_{\theta ^{\pm }_{0}}}\neq 0\]
 where the first term on the left hand side is evaluated for the both limits
\( \theta \rightarrow  \)\( \theta _{0}^{\pm } \).

To see these results, we shall find it convenient to make use of the earlier
discussion in Section 3. We find from (\ref{5.12})\begin{equation}
\label{aaa}
I(p,m,\theta ,\varepsilon )=\frac{A}{\theta }\int ^{1}_{0}dx\ln \frac{[m^{2}-x(1-x)(\theta ^{2}p^{2}_{0}-|\mathbf{p}|^{2})-i\varepsilon ]}{\Lambda ^{2}}
\end{equation}
 We are presently interested in discussing the evaluation of the \( \theta  \)-derivative
and also as to what happens in the two different orders of the limit and the
derivative. So we will drop the term \textasciitilde{} \( \ln \Lambda ^{2} \)
and focus attention on the finite part \( \frac{A}{\theta }I'(p,m,\theta ,\varepsilon ) \).
We note that the logarithm has a phase varying from -\( \pi  \) to \( \pi  \)
for \( \Re es'>4m^{2} \). We then have, using (\ref{5.13}),\begin{eqnarray}
I'(p,m,\theta ,\varepsilon ) & = & \int ^{1}_{0}dx\frac{1}{2}\ln \{[m^{2}-x(1-x)(\theta ^{2}p^{2}_{0}-|\mathbf{p}|^{2})]^{2}+\varepsilon ^{2}\}\nonumber \\
 &  & +\int ^{1}_{0}dx[-\pi i+i\arctan \{\frac{\varepsilon }{x(1-x)(\theta ^{2}p^{2}_{0}-|\mathbf{p}|^{2})-m^{2}}\}]\label{6.3} 
\end{eqnarray}

We shall now focus attention on the imaginary part of \( I' \). We have,\begin{equation}
\label{6.4}
ImI'=\int ^{1}_{0}dx\{-\pi +\arctan \frac{\varepsilon }{B}\}
\end{equation}
 where, as defined in Section 4,\begin{equation}
\label{6.5}
B(\alpha ,\theta ,p)\equiv [-m^{2}+x(1-x)(\theta ^{2}p^{2}_{0}-|\mathbf{p}|^{2})]=[-m^{2}+x(1-x)s']
\end{equation}
 Now,\begin{equation}
\label{6.5a}
\begin{array}{c}
lim\\
\varepsilon \rightarrow 0
\end{array}ImI'=-\int ^{1}_{0}dx\pi \Theta (B)
\end{equation}
 We, thus, find\begin{equation}
\label{6.8}
\frac{\partial }{\partial \theta }\begin{array}{c}
lim\\
\varepsilon \rightarrow 0
\end{array}ImI'=-\pi \int ^{1}_{0}dx\delta (B)\frac{\partial B}{\partial \theta }
\end{equation}
 We note\begin{equation}
\label{6.9}
\delta (B)=\delta (-y^{2}s'+(\frac{1}{4}s'-m^{2}))=\frac{1}{|s'|}\delta (y^{2}-a^{2})
\end{equation}
 with \( y=x-\frac{1}{2} \); and\begin{equation}
\label{6.10}
a^{2}=\frac{s'-4m^{2}}{4s'};\quad a\equiv +\frac{\sqrt{s'-4m^{2}}}{2\sqrt{s'}}
\end{equation}
 For \( s'<4m^{2} \), \( a^{2}<0 \) and the \( \delta  \)-function does not
contribute to the integrand. For \( s'>4m^{2} \), we use\begin{equation}
\label{6.11}
\delta (y^{2}-a^{2})=\frac{1}{2a}\{\delta (y-a)+\delta (y+a)\}
\end{equation}
 and \begin{equation}
\label{6.12}
B=(\frac{1}{4}-y^{2})s'-m^{2}
\end{equation}

\begin{equation}
\label{6.13}
\frac{\partial B}{\partial \theta }=2\theta p^{2}_{0}(\frac{1}{4}-y^{2})
\end{equation}
 to find\begin{equation}
\label{6.14}
\frac{\partial }{\partial \theta }ImI'=-\pi \int ^{1}_{0}dx\delta (B)\frac{\partial B}{\partial \theta }=\frac{-2\pi \theta p^{2}_{0}}{2as'}\int dy(\frac{1}{4}-y^{2})\{\delta (y-a)+\delta (y+a)\}
\end{equation}
 i.e.\begin{equation}
\label{6.15}
\frac{\partial }{\partial \theta }\begin{array}{c}
lim\\
\varepsilon \rightarrow 0
\end{array}ImI'==\frac{-2\pi \theta p^{2}_{0}}{\sqrt{s'-4m^{2}}\sqrt{s'}}(\frac{1}{2}-2a^{2})
\end{equation}
 We thus find that as \( \theta \rightarrow \theta _{0}, \) where \( s'[\theta _{0}]=4m^{2} \),
\( a\rightarrow 0 \) and\begin{equation}
\label{6.16}
\begin{array}{c}
lim\\
\theta \rightarrow \theta ^{+}_{0}
\end{array}\frac{\partial }{\partial \theta }\begin{array}{c}
lim\\
\varepsilon \rightarrow 0
\end{array}ImI'\rightarrow \infty 
\end{equation}
 On the other hand, \begin{equation}
\label{6.17}
\begin{array}{c}
lim\\
\theta \rightarrow \theta ^{-}_{0}
\end{array}\frac{\partial }{\partial \theta }\begin{array}{c}
lim\\
\varepsilon \rightarrow 0
\end{array}ImI'=0
\end{equation}
 Thus, the imaginary part of \( I' \) is not differentiable at \( \theta =\theta _{0}. \)Furthermore,
we note that for a given \( \epsilon >0 \),\begin{equation}
\label{6.18}
\left\{ \begin{array}{c}
lim\\
\theta \rightarrow \theta ^{+}_{0}
\end{array}\frac{\partial }{\partial \theta }ImI'-\begin{array}{c}
lim\\
\theta \rightarrow \theta ^{-}_{0}
\end{array}\frac{\partial }{\partial \theta }ImI'\right\} =0
\end{equation}
 with each given by\begin{equation}
\label{6.18a}
\begin{array}{c}
lim\\
\theta \rightarrow \theta ^{+}_{0}
\end{array}\frac{\partial }{\partial \theta }ImI'=\begin{array}{c}
lim\\
\theta \rightarrow \theta ^{-}_{0}
\end{array}\frac{\partial }{\partial \theta }ImI'=\int ^{1}_{0}dx\{-\frac{\varepsilon }{B^{2}+\varepsilon ^{2}}\frac{\partial B}{\partial \theta }\}\Vert _{_{\theta _{0}}}
\end{equation}
 Thus,\begin{equation}
\label{6.19}
\begin{array}{c}
lim\\
\varepsilon \rightarrow 0
\end{array}\left\{ \begin{array}{c}
lim\\
\theta \rightarrow \theta ^{+}_{0}
\end{array}\frac{\partial }{\partial \theta }ImI'-\begin{array}{c}
lim\\
\theta \rightarrow \theta ^{-}_{0}
\end{array}\frac{\partial }{\partial \theta }ImI'\right\} =0
\end{equation}

Finally, we shall give the treatment for \( m^{2}=0 \) exactly, which differs
somewhat from the above. In this case, \( B \) of (\ref{6.5}) becomes,

\begin{equation}
\label{6.20}
B(\alpha ,\theta ,p)\equiv x(1-x)(\theta ^{2}p^{2}_{0}-|\mathbf{p}|^{2})=x(1-x)s'
\end{equation}
 and thus (\ref{6.5a}) is modified to,\begin{eqnarray}
\begin{array}{c}
lim\\
\varepsilon \rightarrow 0
\end{array}ImI' & = & -\int ^{1}_{0}dx\pi \Theta (s')\nonumber \\
 & = & -\pi \Theta (\theta ^{2}p^{2}_{0}-|\mathbf{p}|^{2})=-\pi \Theta (\theta -\frac{|\mathbf{p}|}{p_{0}})\label{6.21} 
\end{eqnarray}
 and thus, (\ref{6.15}) is modified to (note: we assume \( 0\leq \theta \leq 1 \)),\begin{equation}
\label{6.8}
\frac{\partial }{\partial \theta }\begin{array}{c}
lim\\
\varepsilon \rightarrow 0
\end{array}ImI'=-\pi \delta (\theta -\frac{|\mathbf{p}|}{p_{0}})
\end{equation}

\section{Extension to a Green's function}

In view of the fact that we have considered only a \emph{part} of a contribution
to \emph{one} of the diagrams for the 4-point function, to illustrate the point,
one might immediately suspect that the problem could disappear in the entire
contribution to the S-matrix element. We note first that WT-identities for gauge
variations are equations written down for \emph{Green's functions} and not just
the S-matrix elements. Moreover, what we wish to drive at at the moment is that
the \emph{path-integral itself is not a differentiable function of \( \theta  \)}
as a result of the problems mentioned in the past sections {[} for a further
elaboration, please see section 11.1{]}. For this purpose, it proves sufficient
if the problem with differentiability persists at least at the level of Green's
functions, even if it were to vanish from S-matrix elements from possible mutual
cancellations between diagrams. In this section, we wish to make several remarks
in this connection.

We first show, in an obvious way, that this problem with differentiabilty persists
for a contribution to an \emph{off-shell} Green's function of the type considered
earlier in section 6. An inspection of the diagrams contributing to the Green's
functions will then make it seem extremely unlikely that \emph{at the level
of off-shell Green's functions} there could persist a cancellation; even if
there were one for S-matrix elements. {[}We have verified this point for the
simpler example outlined in the Appendix A{]}.

To see this, we recall that this diagram depends only on \( s'=\theta ^{2}p^{2}_{0}-|\mathbf{p}|^{2} \)
and not on individual 4-momenta \( p_{1,}p_{2} \)and \( p_{3} \). We could
thus choose any \emph{off-shell} \( p_{1} \)and \( p_{2} \) consistent with
the condition that \( s \) is real with \( s>4m^{2} \) and \( \theta  \)\( _{0} \)
is such that \( s'=\theta _{0}^{2}p^{2}_{0}-|\mathbf{p}|^{2}=4m^{2} \). Then
for all such \emph{off-shell} choices of \( p_{1} \)and \( p_{2} \), the above
problem with the differentiability will persist for this contribution.

In appendix A, we shall consider the example of a scalar QED and consider the
off-shell 1-loop Green's function for the process \( \phi \phi \rightarrow \phi \phi  \).
We have verified that the imaginary part of such a process {[}with some further
restrictions on momenta, spelt out in the appendix{]} does have a discontinuous
behaviour on account of the threshold due to unphysical photons and moreover
the threshold for these does depends on \( \theta  \). These are the essential
ingredients that lead to non-differentiabilty.{]}

\section{A Close Analysis of the Example in Section 6}

In this section, we shall qualitatively try to understand why the derivative
with respect to \( \theta  \) does not exist at some point and why the order
of differentiation \( \frac{\partial }{\partial \theta } \) and the limit\( \begin{array}{c}
lim\\
\varepsilon \rightarrow 0
\end{array} \)makes a significant difference there. We can understand this with the help of
a close look at the analyticity properties of the integral \( I'(p,m,\theta ,\varepsilon ) \)
in the Section 6. We recall \begin{equation}
\label{8.1}
I'(p,m,\theta ,\varepsilon )=\int dx\ln \{m^{2}-x(1-x)(\theta ^{2}p^{2}_{0}-|\mathbf{p}|^{2})-i\varepsilon \}
\end{equation}
 The analyticity properties of the integrand above, for a given \( p \) depend
upon \( x \) and \( \theta  \). For a real \( s=p^{2}<4m^{2} \) and a \( \theta \in  \){[}0,1{]},
\( s'=p'^{2}<4m^{2} \) also; and hence for all \( x \) \( \in  \){[}0,1{]}
and all \( \theta \in  \){[}0,1{]}, the integrand is analytic even as \( \varepsilon \rightarrow 0 \).
A small {[}enough{]} variation of \( \theta  \) will not alter these facts
and hence there is no problem occurring in the order of limits \( \delta  \)\( \theta  \)\( \rightarrow 0 \)
and \( \varepsilon \rightarrow 0 \) in this unphysical region. (Besides, in
the unphysical region, the entire calculation can be done in the Euclidean field
theory which would not need \( \epsilon  \)).

We shall, of course, be interested in this issue in the \emph{physical} region
\( s=p^{2}>4m^{2} \). For a given \( s>4m^{2} \), there is a range of \( \theta \in [0,1] \),
such that there \( s'>4m^{2} \) and a range of \( x\in [0,1] \) exists for
which the integrand is close to the branch-point. Also, for any \( \theta \in [0,1] \)
, \( s'>4m^{2} \) \emph{for all \( s \) greater than a certain lower bound}
( See section 11 for more details). Thus, this issue of analytic properties
cannot be avoided for any physical region if one is to interpolate between the
Feynman and the Coulomb gauge.

Next, we note that for a fixed \( p \), the variables \( s' \)and \( \theta  \)
are related: a small variation in \( \theta  \) induces a small variation in
\( s' \) which is the variable relevant to the analytic properties of the integrand.
Thus, for example, we choose a value of \( \theta =\theta _{0} \) such that
\( s' \) is slightly smaller than \( 4m^{2} \), then for a range of \( x \)
\( \varepsilon [0,1] \) this point is very close to the branch-point of the
integrand in the complex \( s' \)-plane. Then the Taylor expansion of the integrand
in \( \theta -\theta _{0} \) is then valid only with a small radius of convergence
\emph{that is directly dependent {[}proportional{]} to \( \epsilon  \)!} In
fact, (\ref{8.1}) will show that the entire integral cannot be expanded in
such a Taylor series except for small\footnote{%
We note that if we took any \( \theta >\theta _{0} \) , the integrand is close
to the branch-cut for a range of \( x \). It may appear that the integral should
then be problematic for all such \( \theta  \) and not just for \( \theta =\theta _{0} \).
This is not the case, at least in this example. We thus end up with just one
troublesome point for \emph{this} diagram that however varies with Lorentz frame
( for this, see section 11). 
} enough \( (\theta -\theta _{0})\sim \varepsilon  \).

We thus see that the fact that the loop-integrand could not be Taylor expanded
for a large enough (\( \theta -\theta _{0} \)) in a \emph{part} of the 4-dimensional
space has, in fact, a reflection on the entire integral. This was anticipated
in Section 3.

The above demonstration now shows why the order of the two limits discussed
earlier matters near \( \theta =\theta _{0} \). If we were to keep \( \epsilon  \)
fixed then there is a small enough radius (\( \theta -\theta _{0} \)) for which
the Taylor expansion holds and the limit (\( \theta -\theta _{0} \))\( \rightarrow 0 \)
can in fact be taken. This procedure can give ultimately

\( \begin{array}{c}
lim\\
\varepsilon \rightarrow 0
\end{array} \)\( \frac{\partial }{\partial \theta } \)\( I(p_{i},\theta ,\varepsilon ) \)

On the other hand, if we are to let \( \epsilon  \)\( \rightarrow 0 \) first,
then the radius of convergence of the Taylor expansion in \( \theta  \) itself
shrinks to zero and we cannot know how to obtain \( \frac{\partial }{\partial \theta } \)\( \begin{array}{c}
lim\\
\varepsilon \rightarrow 0
\end{array} \)\( I(p_{i},\theta ,\varepsilon ) \) by differentiation of the integrand and
subsequent use of the WT identities. Evaluation of the gauge dependence requires
us to evaluate the latter object.

We shall now comment on the treatment of the field theory using the Wick rotation.
It may be thought that the troubles in section 4 and 6 are artificial because
after all we could carry out the Wick rotation in (\ref{5.7}) and (\ref{6.1})
and thus go to the Euclidean field theory that does not require \( \epsilon  \).
Thus, it may be argued that the problems that depend on the limits involving
\( \epsilon  \) and those that depend on the size of \( \epsilon  \) should
be artificial. Now, the field theory of Section 4 and that used in Section 6
are in fact well behaved with respect to the operation \( \frac{\partial }{\partial \theta } \)
in the unphysical region. To see, how the problem could arise in the physical
region, we recall the fact that \( I' \) is a function of \( s'=\theta ^{2}p^{2}_{0}-|\mathbf{p}|^{2} \)
and it has a branch point at \( s'=4m^{2} \). Of course, for \( \Re es'<4m^{2} \),
the function is analytic function of \( s' \) and therefore a differentiable
function of \( \theta  \) ( with \( p_{\mu } \) fixed). But, as \( \epsilon  \)\( \rightarrow  \)0,
the derivative with respect to \( s' \) and hence with respect to \( \theta  \)
{[}for a fixed \( p_{\mu } \){]} cannot exist at the branch point. Thus, the
fact that the generating functional appears (formally) differentiable in \( \theta  \)
in the Euclidean region is no mystery ; but by no means guarantees good behavior
everywhere in the physical region and that is where we are interested in the
gauge-independence issue.

\section{Relation between the altered propagator structure in {[}5{]} and present results}

In reference \cite{j}, we had discussed the gauges interpolating between the
Coulomb and the Feynman gauge with a simple \( \epsilon  \)-term. We had noted
some unusual features of what happens when the interpolating parameter \( \theta  \),
in the gauge interpolating between the Coulomb and the Feynman gauge, is varied.
We had shown that it was important to pay particular attention to the \( \epsilon  \)-term
in the discussion of the gauge-independence. We had considered how the \( \epsilon  \)-term
should be modified with \( \theta  \) if we are to keep the vacuum expectation
value of a gauge-invariant operator unchanged. We had further shown that as
a result of this modification in the \( \epsilon  \)-term, the free propagator
undergoes a radical change in form as the parameter \( \theta  \) is varied
through\footnote{%
The relation between \( \delta \theta  \) and \( \epsilon  \) given here is
a abbreviated way of expressing the condition between these two quantities of
different dimensions. See \cite{j} for the exact condition. 
} a \( \delta  \)\( \theta  \)>\textcompwordmark{}> \( \epsilon  \). We had
argued, in fact, from this observation that such interpolating gauges that assume
any standard (fixed) \( \epsilon  \)-term cannot preserve gauge-independence
as \( \theta  \) is varied and thus do not interpolate correctly between the
Feynman and the Coulomb gauge; while on the other hand trying to modify the
\( \epsilon  \)-term as required for gauge-independence leads to pathological
behavior in the path-integral.

In this section, we shall make a contact between the concrete results obtained
in this work in Sec. 6 and this result obtained earlier in \cite{j}.

We shall establish a simple result to begin with. We shall then relate it to
the question of the variation of the 4-point function with respect to \( \theta  \)
discussed in section 6. It reads\begin{equation}
\label{8.2}
\begin{array}{c}
lim\\
\varepsilon \rightarrow 0
\end{array}I'(\theta ,CC)=\begin{array}{c}
lim\\
\varepsilon \rightarrow 0
\end{array}I'(\theta ,\frac{1}{2}[C+A]\frac{1}{2}[C+A])-\pi i\int ^{1}_{0}dx\Theta [B]
\end{equation}
 Here we have defined, in all cases employed below,\begin{equation}
\label{8.2a}
I(\theta )=\frac{A}{\theta }I'
\end{equation}
 and further \( I(\theta ,\frac{1}{2}[C+A]\frac{1}{2}[C+A]) \) refers to the
(truncated) fish-diagram amplitude obtained by taking the propagator as \( \frac{1}{2} \)(C+A)
in place of C. Here, C refers to the causal propagator and A to the {}``anti-causal{}''
propagator (anti-causal means the one with \( i\varepsilon \rightarrow -i\varepsilon  \)
in C ). The above relation, as we shall later see from Eq. (\ref{8.8}) , represents
in a different manner, the rapid effect of variation with respect to \( \theta  \)
(i.e. \( \frac{\partial I}{\partial \theta }\mid _{_{\theta ^{+}_{0}}}\rightarrow \infty  \)
) which is represented by a replacement of the propagator {}``C{}'' by {}``\( \frac{1}{2}(C+A) \){}''
.

To prove this result, we consider the fish amplitude \( I \) with propagators
\( \frac{1}{2}(C+A) \). We define respectively \( I(\theta ,CC),\: I(\theta ,CA) \)
as the amplitude in question with both the propagators taken as causal, and
with one causal and one anti-causal . Then\begin{equation}
\label{8.3}
I'(\theta ,CA)=I'(\theta ,AC)
\end{equation}
 and\begin{equation}
\label{8.4}
I'[\theta ,\frac{1}{2}(C+A)\frac{1}{2}(C+A)]=\frac{1}{4}\{I'[\theta ,CC]+I'[\theta ,AA]+2I'[\theta ,CA]\}
\end{equation}
 We can then show that as \( \epsilon  \)\( \rightarrow 0 \),\begin{equation}
\label{a1}
\begin{array}{c}
lim\\
\varepsilon \rightarrow 0
\end{array}\left( I'[\theta ,AA]-I'[\theta ,CC]\right) =2\pi i\int ^{1}_{0}dx\Theta [B]
\end{equation}
\begin{equation}
\label{a2}
\begin{array}{c}
lim\\
\varepsilon \rightarrow 0
\end{array}\left( I'[\theta ,CA]-I'[\theta ,CC]\right) =\pi i\int ^{1}_{0}dx\Theta [B]
\end{equation}
 The relation (\ref{a1}) is understood easily: as one goes from the \( I'[\theta ,AA] \)
to \( I'[\theta ,CC] \) by a change \( \epsilon \rightarrow -\varepsilon  \)
and one picks up the discontinuity across the branch-cut given by \( 2\pi i\int ^{1}_{0}dx\Theta [B] \).
Thus, we then have as \( \epsilon  \)\( \rightarrow 0 \),\begin{equation}
\label{a}
\begin{array}{c}
lim\\
\varepsilon \rightarrow 0
\end{array}I'[\theta ,\frac{1}{2}(C+A)\frac{1}{2}(C+A)]=\begin{array}{c}
lim\\
\varepsilon \rightarrow 0
\end{array}I'[\theta ,CC]+\pi i\int ^{1}_{0}dx\Theta [B]
\end{equation}
 We shall now apply this to the amplitude of (\ref{aaa}). Let \( \theta  \)\( _{0} \)
be such that \[
s'=\theta ^{2}_{0}p^{2}_{0}-|\mathbf{p}|^{2}=4m^{2}\]
 We shall let \( \theta =\theta _{0}+\delta \theta  \). We shall assume that
\( \delta \theta >>\varepsilon  \). To understand the significance of the above
result of Eq. (\ref{a}), in the present context, we compare it with the result
that expresses the discontinuous behavior of \( I \) at \( \theta =\theta _{0} \).
It reads,\begin{equation}
\label{b}
\begin{array}{c}
lim\\
\varepsilon \rightarrow 0
\end{array}I'(\theta ,CC)=\begin{array}{c}
lim\\
\varepsilon \rightarrow 0
\end{array}I'(\theta _{0},CC)-\pi i\int ^{1}_{0}dx\Theta [B]+O[\delta \theta ]
\end{equation}
 A comparison of Equations (\ref{a}) and (\ref{b}) shows that as \( \epsilon  \)\( \rightarrow 0 \),
\begin{equation}
\label{8.8}
\begin{array}{c}
lim\\
\varepsilon \rightarrow 0
\end{array}I'(\theta _{0},CC)=\begin{array}{c}
lim\\
\varepsilon \rightarrow 0
\end{array}I'[\theta ,\frac{1}{2}(C+A)\frac{1}{2}(C+A)]+O[\delta \theta ]
\end{equation}
 The above equation is an alternate way of representing the nonsmooth behavior
of \( I \): As \( \theta  \) is varied from \( \theta  \)\( _{0} \) to \( \theta  \)\( _{0}+\delta \theta  \),
\( I \) varies drastically so that the disproportionate change in \( I' \)
viz. \( \{\begin{array}{c}
lim\\
\varepsilon \rightarrow 0
\end{array}I'(\theta ,CC)-\begin{array}{c}
lim\\
\varepsilon \rightarrow 0
\end{array}I'(\theta _{0},CC)\} \) is alternately expressed as that represented by a drastic change in the propagator
structure from \( C\rightarrow \frac{1}{2}[C+A] \). Then the two quantities
\( I(\theta _{0},CC) \) and \( I[\theta ,\frac{1}{2}(C+A)\frac{1}{2}(C+A)] \),
evaluated with \emph{different propagators} for neighboring \( \theta  \)'s,
differ only by a term of \( O[\delta \theta ] \).

This conclusion can be understood well in the light of the work of reference
\cite{j}. There it was found that when \( \theta \rightarrow \theta +\delta \theta  \)
and \emph{the corresponding the gauge variation of the \( \epsilon  \)-term
is also taken into account}, the propagator changed from \( C\rightarrow \frac{1}{2}[C+A] \)
provided \( \delta \theta >>\varepsilon  \). The above conclusion (\ref{8.8})
is a reflection of this. Normally, the WT-identities are taken to imply that
under \( \theta \rightarrow \theta +\delta \theta  \), the corresponding change
in a Green's function is an infinitesimal of \( O(\delta \theta ) \). The above
example explicitly shows that that need not be so, and the Eq. (\ref{8.8})
further shows that a non-trivial change in the Green's function can be correlated
to the change in the \( \epsilon  \)-term arrived at in \cite{j}. Only after
this {}``large{}'' change has been removed, then the \emph{residue} is of
\( O(\delta \theta ) \). We shall have more to say about this in the next section
where we will compare the result (\ref{8.8}) with that of the carefully evaluated
WT-identity.

\section{Expansion of an exponential with an {}``infinitesimal{}'' exponent}

In this section, we shall establish a further contact of the results in the
earlier sections with the work of reference \cite{j}. In reference \cite{j},
we had established several new observations in the context of the interpolating
gauges such as those considered in earlier sections. The ones relevant here
are:

(a) In the treatment of gauge variation, it was necessary to take into account
the \( \epsilon  \)-term carefully.

(b)The gauge variation of the \( \epsilon  \)-term had to be fully taken into
account \emph{and could not be expanded out as an infinitesimal exponent.}

To establish a contact between the analysis of earlier sections and these results,
we shall now follow a procedure parallel to that in \cite{j}.

We now consider the generating functional \begin{equation}
\label{10.1}
W[J;\theta -\delta \theta ]=\int D\phi \exp \{iS_{eff}[A,c,\overline{c};\theta -\delta \theta ]+\varepsilon R+i\int d^{4}xJ^{\mu }A_{\mu }\}
\end{equation}
 where \( S_{eff} \) refers to the effective action in the interpolating gauge
that interpolates between the Coulomb and the Feynman gauge. We imagine performing
the transformation as in \cite{j},\begin{equation}
A'^{\alpha }_{\mu }(x)-A^{\alpha }_{\mu }(x)\equiv \delta A^{\alpha }_{\mu }(x)=iD_{\mu }^{\alpha \beta }c^{\beta }(x)\int d^{4}z\overline{c}(z)\frac{\partial F^{\gamma }[A(z);\theta ]}{\partial \theta }\mid _{_{\theta }}\delta \theta 
\end{equation}

\begin{equation}
\delta c^{\alpha }(x)=-i\frac{1}{2}gf^{\alpha \beta \gamma }c^{\beta }(x)c^{\gamma }(x)\int d^{4}z\overline{c}(z)\frac{\partial F^{\gamma }[A(z);\theta ]}{\partial \theta }\mid _{_{\theta }}\delta \theta 
\end{equation}
\begin{equation}
\label{10.4c}
\delta \overline{c}^{\alpha }(x)=iF^{\alpha }[A(x);\theta ]\int d^{4}z\overline{c}(z)\frac{\partial F^{\gamma }[A(z);\theta ]}{\partial \theta }\mid _{_{\theta }}\delta \theta 
\end{equation}
 As shown in \cite{j}, to preserve the vacuum-expectation-value of a gauge-invariant
operator under this transformation, it is required that the \( \epsilon  \)-term
is suitably changed from \( \varepsilon R\rightarrow  \)\( \epsilon (R+\delta R) \)
{[} For a definition of \( \delta R \), see Eq.(\ref{21.9}){]}. It is then
easy to show that \cite{j}\begin{eqnarray}
W[J;\theta -\delta \theta ] & = & \int D\phi \exp \{iS_{eff}[A,c,\overline{c};\theta -\delta \theta ]+\varepsilon R+i\int J^{\mu }A_{\mu }d^{4}x\}\\
 & = & \int D\phi '\exp \{iS_{eff}[A',c',\overline{c'};\theta ]+\varepsilon [R+\delta R]\nonumber \\
 &  & +i\int J^{\mu }[A'_{\mu }-\delta A_{\mu }]d^{4}x\}\label{10.5aa} 
\end{eqnarray}
 We now suppress the primes in (\ref{10.5aa}), and rewrite the two equations
as\begin{eqnarray}
 &  & \int D\phi \exp \{iS_{eff}[A,c,\overline{c};\theta -\delta \theta ]+\varepsilon R+i\int J^{\mu }A_{\mu }\}\nonumber \\
 & = & \int D\phi \exp \{iS_{eff}[A,c,\overline{c};\theta ]+\varepsilon [R+\delta R]+i\int d^{4}xJ^{\mu }[A_{\mu }-\delta A_{\mu }]\label{10.5ab} 
\end{eqnarray}
 Or written differently,\begin{equation}
\label{bb}
<<\exp \{\frac{1}{2}i\int d^{4}x\Delta F^{2}-i\Delta S_{gh}\}>>\mid _{_{\theta }}=<<\exp \{\varepsilon \delta R-i\int d^{4}xJ^{\mu }\delta A_{\mu }\}>>\mid _{_{\theta }}
\end{equation}
 where, we have introduced the short-hand notation\begin{equation}
\label{10.5a}
<<O[\phi ]>>\mid _{_{\theta }}\equiv \int D\phi O[\phi ]\exp \{iS_{eff}[A,c,\overline{c};\theta ]+\varepsilon R+i\int d^{4}xJ^{\mu }A_{\mu }\}
\end{equation}
 and \[
\Delta F^{2}\equiv F^{2}(\theta )-F^{2}(\theta -\delta \theta );\; \Delta S_{gh}=S_{gh}(\theta )-S_{gh}(\theta -\delta \theta )\]
 The relation (\ref{bb}) as it stands is undoubtedly correct. In the light
of results of Sections 4 and 6 and of reference \cite{j}, we however ask whether
the simplified version of this result, obtained by expanding the exponential
with an {}``infinitesimal{}'' exponents, viz. \begin{equation}
\label{10.5}
<<\{\frac{1}{2}i\int d^{4}x\Delta F^{2}-i\Delta S_{gh}\}>>\mid _{_{\theta }}=<<\{\varepsilon \delta R-i\int d^{4}xJ^{\mu }\delta A_{\mu }\}>>\mid _{_{\theta }}
\end{equation}
 i.e.\begin{equation}
\label{c}
<<\{\frac{1}{2}i\int d^{4}x\Delta F^{2}-i\Delta S_{gh}-\varepsilon \delta R+i\int d^{4}xJ^{\mu }\delta A_{\mu }\}>>\mid _{_{\theta }}=0
\end{equation}
 and also the one obtained by dropping the \( \varepsilon \delta R \) term
altogether in (\ref{c}), viz.\begin{equation}
\label{c1}
<<\{\frac{1}{2}i\int d^{4}x\Delta F^{2}-i\Delta S_{gh}+i\int d^{4}xJ^{\mu }\delta A_{\mu }\}>>\mid _{_{\theta }}=0
\end{equation}
 (which we normally understand as the WT-identity relevant to the evaluation
of the \( \theta  \)-dependence of Green's functions), can actually be used,
in this form, to evaluate the gauge variation \emph{always}\footnote{%
We note, as mentioned earlier, that the WT-identity of the form (\ref{c}) does
not \emph{always} follow by the procedure of expanding the exponential . Nevertheless,
it could have been arrived at formally by writing down the BRS WT-identity and
acting on it by the functional differential operator \( \delta \theta \int d^{4}z\{\frac{\delta }{\delta \xi (z)} \)
\( \frac{\partial F^{\gamma }}{\partial \theta }[-i\frac{\delta }{\delta J(z)};\theta ]\} \)
. (Here, \( \xi  \) refers to the source of \( \overline{c} \)). However,
such a WT-identity will not enable us evaluate the gauge variation of a Green's
function such as that in Section 6-7, when we want to jump across \( \theta =\theta _{0} \),
since it will not pick up the change in the nondifferentiable part (imaginary
part in the context of the example in Sec. 6-7) correctly. There, we will \emph{have}
to use (\ref{bb}); where the effects of the \( \epsilon  \)-term cannot be
overlooked. Further, the WT-identity (\ref{c}) cannot be {}``exponentiated
back{}'' to yield (\ref{bb}) i\emph{n the neighborhood of such a point.}
}. We consider this issue in light of the results in section 4, where we found
that \begin{equation}
\label{10.6}
<<\exp [i\delta \alpha \int d^{4}x\phi \partial ^{2}_{0}\phi ]-1>>
\end{equation}
 could not always be interpreted as\begin{equation}
\label{10.7}
<<[i\delta \alpha \int d^{4}x\phi \partial ^{2}_{0}\phi ]>>
\end{equation}
 and it was later explicitly understood during the discussion regarding the
example considered in section 6. We had also seen in reference \cite{j} that
the effect of the modification of the \( \epsilon  \)-term was by no means
always infinitesimal: It, in fact, lead to a unexpected modification of the
propagator structure. In the normal usage of the WT-identity {[}2,3{]}, we not
only carry out this expansion, but in fact \emph{ignore} the \( \epsilon  \)\( \delta  \)R
term altogether\footnote{%
For unbroken gauge theories in the Lorentz class of gauges, we believe this
is justified either by noting that no modification of the pole structrure happens
in this case by the \( \epsilon  \)-term variation \( \epsilon  \)\( \delta  \)R
or by independent arguments. 
}.

We shall now try to apply the WT-identity (\ref{bb}) to the special case of
the 4-point function considered in Section 6. We shall look at the left hand
side of (\ref{bb})in this context. To this 4-point function in the 1-loop approximation,
there are a number of diagrams contributing. The effect of diagrams contributing
to this term is to give the change in the 4-point function in the one loop approximation
when the parameter \( \theta  \) is changed from \( \theta  \)-\( \delta  \)\( \theta  \)
to \( \theta  \). Among this difference there is the contribution which is
exactly of the form of the difference\[
\begin{array}{c}
lim\\
\varepsilon \rightarrow 0
\end{array}\{I(\theta )-I(\theta -\delta \theta )\}\]
 As seen, this difference is not understood as \( \delta \theta \frac{dI}{d\theta } \)
for a process at such\footnote{%
For the compatibility of notations with that used earlier in this section, we
have somewhat altered our conventions for \( \delta  \)\( \theta  \). 
} a \( \theta -\delta \theta =\theta _{0} \) that \( s'(\theta _{0})=4m^{2} \).
At this point, it may appear that as we are taking only one of the contributions
to the S-matrix elements, such discontinuities will cancel out when all contributions
are taken into account. Here, we would like to draw attention to the remarks
made in section 7 that such problems are by no means specific to the contribution
to the S-matrix element, they are also present in the off-shell Green's functions
where we \emph{do not} expect fortuitous cancellations of this kind to happen.
{[}This fact has explicitly been verified in the example considered in Appendix
A{]}.

Now, this contribution \( [I(\theta )-I(\theta -\delta \theta )] \) cannot
be expanded as \( \frac{dI}{d\theta }\delta \theta  \); so that it is not possible
to expand out (at least) this contribution to\( <<\exp \{\frac{1}{2}i\int d^{4}x\Delta F^{2}\}-1>>\mid _{_{\theta }} \)
as arising from \( <<\frac{1}{2}i\int d^{4}x\Delta F^{2}\}>>\mid _{_{\theta }} \)as
then it would have to be proportional to \( \delta  \)\( \theta  \). It is
the latter term that is normally taken as one of the terms in the WT-identity
after assuming that such an expansion does indeed hold!

We next move onto the term\( <<\exp \{\varepsilon \delta R\}>>\mid _{_{\theta }} \)on
the right hand side of (\ref{bb}). We would normally assume that we can expand
\( \exp \{\varepsilon \delta R\} \) as\[
\exp \{\varepsilon \delta R\}\simeq 1+\varepsilon \delta R\]
 and we normally drop the term \( \varepsilon \delta R \) on the right hand
side of the WT-identity (\ref{c}). We had however seen an unusual effect of
this term in \cite{j} in the context of the present example of the interpolating
gauge, which was neither infinitesimal nor ignorable ( See section 2 for more
details). In fact, according to \cite{j}, we \emph{cannot} always treat this
term as of a lesser order compared to the original \( \epsilon  \)-term, \( \varepsilon R \),
in the exponent: its effects could be drastic enough to alter the propagator
structure for \( \delta \theta >>\varepsilon  \). (This was essentially because
\( \varepsilon \delta R \) can contribute to an inverse propagator in such
a way that the contribution blows up in a sensitive kinematical region and can
overwhelm the \( \epsilon R \) term itself. The net \( \epsilon  \)-term then
determines the new propagator structure). Thus, it is by no means obvious that
the expansion of \( \exp \{\varepsilon \delta R\} \) can be carried out nor
is it obvious that its effects can be ignored as it is normally done.

We shall explicitly show this in working for the example we have worked out
in section 9. To see this, we restructure Eq.(\ref{8.8}) as follows:\begin{eqnarray}
 &  & \begin{array}{c}
lim\\
\varepsilon \rightarrow 0
\end{array}I'(\theta _{0},CC)-\begin{array}{c}
lim\\
\varepsilon \rightarrow 0
\end{array}I'(\theta ,CC)\nonumber \\
 & = & \left\{ \begin{array}{c}
lim\\
\varepsilon \rightarrow 0
\end{array}I'[\theta ,\frac{1}{2}(C+A)\frac{1}{2}(C+A)]-\begin{array}{c}
lim\\
\varepsilon \rightarrow 0
\end{array}I'(\theta ,CC)\right\} \nonumber \\
 &  & +O[\delta \theta ]\label{10.8} 
\end{eqnarray}

We now compare the above equation with (\ref{bb}). We note that the left hand
side is a particular contribution to \( <<\{\frac{1}{2}i\int d^{4}x\Delta F^{2}\}-1>>\mid _{_{\theta }} \)
from the diagram considered. The curly bracket on the right hand side is the
corresponding ({}``large{}'') contribution from \( <<\exp (\varepsilon \delta R)-1\}>>\mid _{_{\theta }} \)and
has arisen from the seemingly {}``infinitesimal{}'' exponent \( \varepsilon \delta R \)\footnote{%
The contribution from \( <<\exp (\varepsilon \delta R)-1\}>>\mid _{_{\theta }} \)
to a truncated diagram having internal gauge boson lines is obtained as the
difference of the diagram evaluated with \( \frac{1}{2}(C+A) \) as the propagator
and the one with C as the propagator \cite{j}. 
}. .We note that it is this contribution that carries in it the {}``drastic{}''
change in \( I \). \emph{It is the} residue \emph{that now is \( O(\delta \theta ) \)}
and (when such residues are now collected for all contributing diagrams) it
can be identified with the usual infinitesimal change from \( <<i\int d^{4}xJ^{\mu }\delta A_{\mu }\}>>\mid _{_{\theta }} \)that
we normally associate with the gauge-variation of a Green's function via WT-identity.

To summarize, we have written out the rigorous WT-identity (\ref{bb}) as would
follow from the path-integral that takes into account the \( \epsilon  \)-term.
We went on to discuss whether the simplifications one usually makes in it are
always valid in the context of such interpolating gauges.We therefore applied
the WT-identity (\ref{bb}) to the (off-shell and \emph{truncated}) 4-point
function of the gauge theories in 1-loop approximation. We write (\ref{bb})
as\begin{equation}
\label{bbbb}
<<\exp \{\frac{1}{2}i\int d^{4}x\Delta F^{2}-i\Delta S_{gh}\}-\exp \{\varepsilon \delta R\}>>\mid _{_{\theta }}=<<-i\int d^{4}xJ^{\mu }\delta A_{\mu }\}>>\mid _{_{\theta }}
\end{equation}
 Contribution to the left hand side of (\ref{bbbb}) comes from a set of diagrams
in pairs (the first term giving the gauge variation of an original diagram to
the Green's function and the second corresponds to the variation from an appropriate
modification of the propagators \( C\rightarrow \frac{1}{2}(C+A) \) in \emph{that}
diagram). We focused attention on a particular pair of contributions to the
two terms coming from two time-like gluons intermediate state. We showed that
{[} for specific kinematical relations between \( p_{i} \)'s and \( \theta  \){]}
this contribution to neither \( <<\exp \{\frac{1}{2}i\int d^{4}x\Delta F^{2}\}>>\mid _{_{\theta }} \)nor\( <<\exp \{\varepsilon \delta R\}>>\mid _{_{\theta }} \)be
expanded out as \( 1+O(\delta \theta ) \). We further noted that the \( \exp \{\varepsilon \delta R\} \)term
could not be dropped out of the WT-identity. We found further that \emph{when
the difference between these contributions is taken, that difference is O(\( \delta  \)\( \theta  \))
.} We expect a similar result to hold for such pairs of terms arising from different
diagrams and confirm the conclusion presented here for the entire 4-point function.

\section{Further comments}

We shall now add several comments.

\subsection{Extension of difficulties associated with definition of \protect\protect\( \frac{\partial }{\partial \theta }\protect \protect \)}

In section 6, we discussed a specific simple contribution to the 4-point function
and studied its properties as \( \theta  \) is varied. We found that for any
physical amplitude, this contribution was not differentiable with \( \theta  \)
at some value of \( \theta  \). Not only did the derivative not exist, the
left and the right derivatives differ by infinity. Moreover, we showed that
the order of the limits \( \epsilon \rightarrow 0 \) and differentiation \( \frac{\partial }{\partial \theta } \)
mattered at this point . We argued that the same effect also holds for a similar
contribution to an off-shell Green's function. From its appearance, the scope
of this result appears limited. In this section, we shall make a comment that
suggests that the scope of this specific result itself is in fact wider than
stated so far and show that the difficulties encountered in Section 6 are by
no means confined to a particular value of \( \theta  \) given \( s>4m^{2} \).
{[}In addition, of course, we would encounter similar problems with other Green's
functions which we have not dealt with in this work{]}.

We shall give a simple argument to see that the difficulty cannot be confined
to a particular value of \( \theta  \) for a given process. We do this by considering
different Lorentz frames.

We note that for a given \( s \) sufficiently larger than \( 4m^{2} \) (we
shall soon be more specific about how much larger) there exists a Lorentz frame
where \( s'=4m^{2} \) for \emph{any} \( \theta \in (0,1] \). To see this,
we note that the existence of the solution for

\[
s'=\theta ^{2}p^{2}_{0}-|\mathbf{p}|^{2}=4m^{2}\]
 with \[
s=p^{2}_{0}-|\mathbf{p}|^{2}>4m^{2}\]
 simply requires that\[
s-4m^{2}>(1-\theta ^{2})p^{2}_{0}=(1-\theta ^{2})(s+|\mathbf{p}|^{2})\]
 This leads to\[
s>\frac{4m^{2}}{\theta ^{2}}+\frac{(1-\theta ^{2})}{\theta ^{2}}|\mathbf{p}|^{2}\geq \frac{4m^{2}}{\theta ^{2}}\]
 \[
s>\frac{4m^{2}}{\theta ^{2}}\]
 also proves to be a sufficient condition for the existence of a Lorentz frame
with \( s'=4m^{2} \).

Thus, the difficulties we encountered in Section 6 will be encountered in \emph{some
frame} for \emph{any} \( \theta \in (0,1] \) provided \( s \) is sufficiently
large.

We now consider the generating functional for the theory expanded in terms of
the gauge field Green's functions in momentum space as: \begin{equation}
\label{11.11}
Z[J,\theta ]=\sum _{n}\int \prod _{i}d^{4}p_{i}J(p_{i})\delta ^{4}(\sum _{i}p_{i})\widetilde{G_{c}}^{(n)}(p_{1},....p_{n};\theta )
\end{equation}
 Focusing our attention on the \( n=4 \) term, for the present, we note that
given \emph{any} \( \theta \in (0,1) \), there exists a \( \widetilde{G_{c}}^{(4)}(p_{1},....p_{4};\theta ) \)
which is not differentiable there. In fact, this is true for all \( \widetilde{G_{c}}^{(4)}(p_{1},....p_{4};\theta ) \)
for which \( p_{1},p_{2},p_{3} \) lie in a certain volume in an 11-dimensional
subspace of the 12-dimensional momentum space. This makes \( Z[J,\theta ] \)
nondifferentiable\footnote{%
Here, we are assuming that we can use sources \( J(p_{i}) \) that can create
sharply defined momentum states (plane wave states). We can also relax this
assumption and use only wavepackets for external lines. For this, please refer
to the following subsection 11.2 . 
} everywhere in \( \theta \in (0,1) \).

\subsection{Wave-packets }

If one does not use sharply defined external line momenta but rather wave-packets,
the value of \( s \) and also \( s' \)for the given diagram in Section 6 are
{}``smeared{}''. Then, with appropriate choice of wave-packets, it is possible
to allow for the definition of \( \frac{\partial }{\partial \theta }I'\mid _{_{smeared}} \)
even around the region of the unphysical particle threshold. However, this is
no more helpful than making \( \frac{\partial }{\partial \theta }I' \) finite
by keeping \( \epsilon >0 \) till the end ( See section 5). In either case,
the discussion in section 10 remains valid: the \( \epsilon  \)-term continues
to contribute the same way except that both the terms on the left hand side
of (\ref{bbbb}) are now simultaneously smeared. The important conclusions are
unaltered.

\section{Conclusions}

We shall now summarize our conclusions.We pointed out that interpolating gauges
necessarily contain parameter-dependent denominators. We considered the special
case of the interpolating gauge of Doust that interpolates between the Feynman
and the Coulomb gauge. We drew attention to the fact that the path-integrals
in such gauges do lead to Green's functions that are not differentiable functions
of the variable parameter \( \theta  \). We dealt with a specific contribution
to a 4-point function in 1-loop approximation in detail. In connection with
this, we established several results. We found that this was not differentiable
at some value of the interpolating parameter \( \theta \in (0,1) \). In fact,
we showed that the path-integral that generates such diagrams has this obstruction
at \emph{every} \( \theta \in (0,1) \). We further showed that this amplitude
at such a \( \theta =\theta _{0} \) is not a differentiable function of \( \theta  \),
and consequently, the gauge variation around this point is not of \( O(\delta \theta ) \)
as the parameter is varied from \( \theta _{0}\rightarrow \theta _{0}+\delta \theta  \).
This contradicts the assumption one makes in the derivation of the WT-identity.
We further found that in the neighborhood of such point, \emph{it was necessary
to keep the contribution from the variation of the \( \epsilon  \)-term} in
the WT-identity. Both of these contributions were {}`` large{}'' in the neighborhood
of this point. We further made the connection with the results of Ref.\cite{j}.

\section{Appendix A}

In this appendix, we shall give an explicit example to substantiate the claims
in Section 7 for the off-shell Green's functions. It will prove simpler to deal
with an abelian gauge theory: we shall consider scalar electrodynamics given
by the Lagrangian density,

\[
\mathcal{L}=(\partial _{\mu }\phi ^{*}+ieA_{\mu }\phi ^{*})(\partial ^{\mu }\phi -ieA_{\mu }\phi )-m^{2}\phi ^{*}\phi -\frac{1}{2}F[A,\theta ]^{2}-\frac{1}{4}F_{\mu \nu }F^{\mu \nu }\]
 For future use, we note the propagator for the gauge boson\cite{d}:\[
G_{\mu \nu }=G^{tr}_{\mu \nu }+G^{T}_{\mu \nu }+G_{\mu \nu }^{L}\]
 with\[
G_{\mu \nu }^{tr}=\frac{ig_{\mu i}g_{\nu j}}{k^{2}+i\varepsilon }(\delta _{ij}-\frac{k_{i}k_{j}}{|\mathbf{k}|^{2}})\]
 \[
G^{T}_{\mu \nu }=-\frac{ig_{\mu 0}g_{\nu 0}}{(\theta ^{2}k^{2}_{0}-|\mathbf{k}|^{2}+i\varepsilon )}\]
 \[
G^{L}_{\mu \nu }=\theta ^{2}\frac{ig_{\mu i}g_{\nu j}k^{i}k^{j}}{(\theta ^{2}k^{2}_{0}-|\mathbf{k}|^{2}+i\varepsilon )}\frac{1}{|\mathbf{k}|^{2}}\]
 are respectively, the transverse, the time-like, and the longitudinal propagators.
We note that both the \( G^{T}_{\mu \nu },G_{\mu \nu }^{L} \) have the same
pole structure, while the transverse part has only a usual pole at \( k^{2}+i\varepsilon =0 \).

We consider the 1-loop contributions to the process:

\[
\phi ^{+}\phi ^{-}\rightarrow \phi ^{+}\phi ^{-}\]
 through 2-photon exchange. We let the 4-momenta of the 4 particles be \( p_{i};i=1,2,3,4 \)
respectively.We shall not necessarily require these to be on-shell but later
choose them suitably.

We consider the imaginary part of the forward amplitude for this process with
\( p_{i}^{2}\leq m^{2} \)and \( s>0 \) and \( u<0 \). In this case, the contribution
to the imaginary part comes only from the two photons in the intermediate states
and there are three possibilities:(1) both photon propagators are \( G^{tr} \);(2)
One photon propagator is \( G^{tr} \)and the other is \( G^{(T,L)} \); and
(3)both the photon propagators are \( G^{(T,L)} \). It turns out that the threshold
for the cases (1) and (2) are at \( (p_{1}+p_{2})^{2}=(k_{1}+k_{2})^{2}=0 \)
and thus are independent of \( \theta  \). Only the threshold for the third
case is at \( \theta ^{2}(p_{10}+p_{20})^{2}-|\mathbf{p}_{1}+\mathbf{p}_{2}|^{2}=0 \)
as also found in Section 6. This \( \theta  \)-dependent threshold gives trouble
with differentiation.

We have verified that the imaginary part arising from the case (3) vanishes
for the on-shell amplitude in the one-loop approximation; which is guaranteed
here from the tree WT-identities (\emph{terms in} \emph{which do not yet depend
on the gauge)}. We have also verified that the imaginary part from the case
(3) does not vanish near threshold for the off-shell amplitude.

\textbf{\emph{ACKNOWLEDGMENTS}}

I would like to acknowledge financial support from Department of Science and
Technology, Government of India in the form of a grant for the project No. DST-PHY-19990170.

1{]} On page 29, in the second line after Eq. (105), replace the part of the
sentence: {}`` actually holds in this form.{}'' by the following (together
with the footnote there):

{}``can actually be used, in this form, to evaluate the gauge variation \emph{always}\footnote{%
We note, as mentioned earlier, that the WT-identity of the form (104) does not
\emph{always} follow by the procedure of expanding the exponential . However,
it could have been arrived at formally by writing down the BRS WT-identity and
acting on it by the functional differential operator \( \delta \theta \int d^{4}z\{\frac{\delta }{\delta \xi (z)} \)
\( \frac{\partial F^{\gamma }}{\partial \theta }[-i\frac{\delta }{\delta J(z)};\theta ]\} \)
. (Here, \( \xi  \) refers to the source of \( \overline{c} \)). However,
such a WT-identity will not enable us evaluate the gauge variation of a Green's
function such as that in Section 6-7, when we want to jump across \( \theta =\theta _{0} \),
since it will not pick up the change in the nondifferentiable part (imaginary
part in the context of the example in Sec. 6-7) correctly. There, we will \emph{have}
to use (101); where the effects of the \( \epsilon  \)-term cannot be overlooked.
Further, the WT-identity (104) cannot be {}``exponentiated back{}'' to yield
(101) i\emph{n the neighborhood of such a point.}
}\emph{.}

\end{document}